\documentclass[a4paper]{article}
\usepackage[margin=3.5cm]{geometry}

\usepackage{amsmath}
\usepackage{amssymb}
\usepackage{mathrsfs}

\usepackage{graphicx}
\usepackage{grffile} 
\usepackage{authblk}
\usepackage{float}
\usepackage{placeins}
\usepackage{color}
\usepackage{dsfont}
\usepackage{listings}

\usepackage[utf8]{inputenc} 

\usepackage{amsthm}

\newtheorem{proposition}{Proposition}

\numberwithin{theorem}{section} % important bit

\def\lf{\left\lfloor}   
\def\rf{\right\rfloor}

\begin{document}

\title{From Rough to Multifractal volatility: the log S-fBM model}
\author[1,*]{Peng Wu}
\author[2]{Jean-François Muzy}
\author[1]{Emmanuel Bacry}

\affil[1]{Ceremade, CNRS-UMR 7534, Université Paris-Dauphine PSL \newline
Place du Maréchal de Lattre de Tassigny, 75016 Paris, France}

\affil[2]{SPE CNRS-UMR 6134, Université de Corse \newline
BP 52, 20250 Corte, France}

\affil[*]{Corresponding author, peng.wu@dauphine.eu}
{
    \makeatletter
    \renewcommand\AB@affilsepx{: \protect\Affilfont}
    \makeatother

    \affil[ ]{Emails}

    \makeatletter
    \renewcommand\AB@affilsepx{, \protect\Affilfont}
    \makeatother

    \affil[2]{muzy@univ-corse.fr}
    \affil[1]{bacry@ceremade.dauphine.fr}
}

\newcommand{\R}{\ensuremath{\mathbb{R}}}
\newcommand{\Q}{\ensuremath{\mathbb{Q}}}
\newcommand{\N}{\ensuremath{\mathbb{N}}}
\newcommand{\Z}{\ensuremath{\mathbb{Z}}}
\newcommand{\Prob}{\ensuremath{\mathbb{P}}}
\newcommand{\ProbQ}{\ensuremath{\mathbb{Q}}}
\newcommand{\bM}{\bar{M}}
\newcommand{\tM}{\widetilde{M}}
\newcommand{\hM}{\widehat{M}}

\newcommand{\E}{\mathbb{E}}
\newcommand{\Var}{\mathrm{Var}}
\newcommand{\Cov}{\mathrm{Cov}}

\newcommand{\biggamma}{\makebox{\large\ensuremath{\gamma}}}

\newcommand{\F}{\mathcal{F}}
\newcommand{\I}{\mathcal{I}}

\renewcommand{\d}{\, \mathrm{d}}
\newcommand*{\dif}{\mathop{}\!\mathrm{d}}
\newcommand\eql{\stackrel{\lambda}{=}}

\providecommand{\keywords}[1]
{
  \small    
  \textbf{\textit{Keywords---}} #1
}

%customised symbols
\newcommand{\ixa}{\ell}
\newcommand{\ixb}{m}

\maketitle

\begin{abstract}
We introduce a family of random measures $M_{H,T} (\d t)$, namely log S-fBM,  such that, for $H>0$, $M_{H,T}(\d t) = e^{\omega_{H,T}(t)} \d t$ where
$\omega_{H,T}(t)$ is a Gaussian process that can be considered as a stationary version of an 
$H$-fractional Brownian motion. Moreover,  when $H \to 0$, one has $M_{H,T}(\d t) \to \tM_{T}(\d t)$ (in the weak sense) where $\tM_{T}(\d t)$ is the celebrated log-normal multifractal random measure (MRM). Thus, this model allows us to consider, within the same framework, the two popular classes of multifractal ($H = 0$) and rough volatility ($0<H < 1/2$) models.
The main properties of the log S-fBM are discussed and their estimation issues are addressed.
We notably show that the direct estimation of $H$ from the scaling properties of $\ln(M_{H,T}([t, t+\tau]))$, at fixed $\tau$, can lead to strongly over-estimating the value of $H$. We propose a better GMM estimation method which is shown to be valid in the high-frequency asymptotic regime.
When applied to a large set of empirical volatility data, we observe that stock indices have values around $H=0.1$
while individual stocks are characterized by values of $H$ that can be very close to $0$ and thus well described by a MRM. We also bring evidence that unlike the log-volatility variance $\nu^2$ whose estimation appears to be poorly reliable (though used widely in the rough volatility literature), the estimation of the so-called "intermittency coefficient" $\lambda^2$, which is the product of  $\nu^2$ and the Hurst exponent $H$, appears to be far more reliable leading to values that seem to be universal for respectively all individual stocks and all stock indices.
\end{abstract} 

%\hspace{10pt} 

\keywords{Rough volatility, Multifractal volatility, fractional Brownian motion, GMM estimation, Intermittency coefficient}

%%%%%%%%%%%%%%%%%%%%%%%%%%%%%%%%%%%%%%%%%%%%%%%%%%%%%%%%%%%%%%%%%%%%%%%
%%%%%%%%%%%%%%%%%%%%%%%%%%%%%%%%%%%%%%%%%%%%%%%%%%%%%%%%%%%%%%%%%%%%%%%
%%%%%%%%%%%%%%%%%%%%%%%%%%%%%%%%%%%%%%%%%%%%%%%%%%%%%%%%%%%%%%%%%%%%%%%

\section{Introduction}

During the past few years, new insights on stochastic volatility models have been obtained after the observation by Gatheral et al. \cite{vol_is_rough} (see also \cite{Bennedsen_2020,Bennedsen2021decoupling}), that the logarithm of the realized volatility is rough, i.e., is less regular than a standard Brownian motion. Rough volatility models have become very popular not only because they allow one to account for main empirical realized volatility properties but also because, when they are considered in asset price models,  they provide a very good fit of option prices and notably their ATM skew power-law behavior close to maturity \cite{rough_vol_evd_option,Bayer_Friz_Gatheral_rBergomi,fuka20}.
The first empirical evidence reported in \cite{vol_is_rough} suggests that the logarithm of the asset price stochastic variance can be represented by a fractional Brownian motion (fBM) of Hurst exponent $H$ close to $H \simeq 0.1 < 1/2$. 
More recent studies based either on quasi-likelihood approach \cite{fuka19} or GMM-approach \cite{rough_gmm}, consistently suggest the $H$ is even closer to $H=0$, i.e., $H \lesssim 0.05 $ for a large panel of equity data.
In that respect, it is natural to consider the limit $H \to 0$ in the rough process driving the volatility logarithm. Even if one cannot plug $H=0$ in the power-law expression of the fractional Brownian motion covariance, formally, it corresponds to a logarithmic behavior. 

Such a logarithmic behavior is precisely the one that characterises the so-called continuous random cascade models introduced two decades ago by Bacry et al. \cite{MDB00,BDM01}. Indeed, in 2000, these authors proposed the ``Multifractal Random Walk" (MRW) as a model for asset prices in order to account for their multifractal properties, i.e., the fact observed by various authors (see e.g. \cite{TF1,TF2,MCF97}) that asset return empirical moments obey non-trivial scaling properties. The MRW model relies on a multifractal stochastic volatility model, namely the  ``Multifractal Random Measure'' (MRM) model
\cite{MDB00,BDM01}, in which the log-volatility is provided by a log-correlated Gaussian field. Such a class of processes, also referred to as Gaussian multiplicative chaos, has been at the heart of many studies in a large variety of applications \cite{RV1}. Gaussian multiplicative chaos and the associated log-normal random cascades have been extended to any infinitely divisible distribution by Bacry and Muzy in \cite{Muzy_2002,bacry_muzy_2003}.

Recovering a multifractal volatility model as the limit $H \to 0$ of a rough volatility model or, from a more general perspective, defining a meaningful limit $H \to 0$ of a fractional Brownian motion and one of its variants has been the subject of various recent studies.  In \cite{Fyod16}, the authors build an $H \! \! =  \! 0$ - fBM by considering a regularisation from the harmonizable representation of fBM's while in \cite{neuman_rosenbaum_2018,Hager_Neuman_zero_Hurst} a $H=0$ limiting process is obtained using a peculiar normalisation and centering of the fBM. In \cite{forde2020} (see also \cite{forde2020roughBergomi}), the authors consider the limit $H \to 0$ of the exponential of a rescaled Riemann-Liouville fBM and its relationship with Gaussian multiplicative chaos. Finally, in \cite{Bayer20}, Bayer et al. propose a new class of rough models that consists in modulating the Riemann-Liouville fBM power-law kernel by a logarithmic factor. The so-obtained "super-rough" stochastic volatility remains well-defined as a continuous process when $H = 0$.

In this paper, our goal is to add a contribution to this problem by introducing a new version of rough volatility models based on the so-called ``stationary'' fBM (S-fBM). S-fBM is a variant of fBM whose covariance function is exactly the one obtained when considering the small-time approximation of the correlation of the fractional Ornstein-Uhlenbeck process considered in \cite{vol_is_rough}. We prove that when $H \to 0$, one recovers the exact self-similar multifractal measure defined in \cite{MDB00,BDM01}. Our construction is based on the same approach proposed in \cite{Muzy_2002,bacry_muzy_2003} where the log-volatility is obtained from the integration of a 2D Gaussian white noise over a triangular domain in a time-scale plane. It turns out such an approach corresponds to the same method defined by Takenaka to build the fractional Brownian motion \cite{Taqqu}.
Our model therefore provides a unified framework to consider both rough and multifractal stochastic volatility
models. Beyond defining the main statistical properties of the model, we aim at estimating its parameters on a large panel of market data. For that purpose, we extend the
GMM method proposed in \cite{bacry_kozhemyak_muzy_2013} that is based on a ``small intermittency" expansion of the moments of the measure logarithms.

The paper is organized as follows: in section \ref{sec:defs}, after recalling the basic notions underlying usual rough volatility models and the definition of the multifractal random measure (MRM), we introduce the log S-fBM random measure $M_{H,T}(\d t)$ as the exponential of the S-fBM random process which is nothing but a ``stationary'' version of the fractional Brownian motion of Hurst parameter $0<H<1$. We show that one 
recovers the celebrated Mandelbrot-Van Ness fBM when $T$, the correlation parameter of our model, tends to infinity. In this section, we also show that the log S-fBM converges, when $H \to 0$, towards a Multifractal Random Measure, consequently leading to a unified framework for rough volatility models ($M_{H,T}(\d t)$, for $H\neq 0$) and multifractal volatility models (by extension, $M_{H=0,T}(\d t)$).
In section  \ref{sec:second_order}, we establish, within this unified framework,  analytical expressions for the second-order moments of respectively $M_{H,T}([t,t+\Delta])$ and its logarithm, while in section \ref{sec:gmm_estimation} we define two GMM parameter estimation methods based on these expressions. Our approach is illustrated by various numerical examples. Application to empirical data, namely the daily volatility of many individual stocks as well as market indices is provided in section \ref{sec:emp}.
Section \ref{sec:conc} summaries our findings while technical material and mathematical proofs are provided in Appendices.

\section{The log Stationary fractional Brownian Motion (log S-fBM) stochastic volatility model}
\label{sec:defs}
\subsection{Multifractal and rough volatility models}

Before introducing our new model of stochastic volatility measure (log S-fBM), let us briefly walk through
the two popular former classes of stochastic volatility models it is notably designed to unify, namely
the Rough Fractional Stochastic Volatility (RFSV) model and the Multifractal Random Walk (MRW) or Multifractal Random Measure (MRM) models.

\paragraph{The MRM/MRW models} The MRW was firstly introduced in 2001 by Bacry et al. \cite{MDB00,BDM01} as a model for log-prices $X(t)$ that has exact (log-normal) multifractal properties, i.e., such 
that the moment of price returns $\delta_\tau X(t) = X(t+\tau)-X(t)$ obeys 
exact scaling properties:
\begin{equation}
\label{scaling_MRW}
  \E[|\delta_\tau X(t)|^q] = \E[|X(t+\tau) -X(t)|^q]\sim C_q \tau^{\zeta(\frac{q}{2})}
\end{equation}
where the multifractal scaling spectrum $\zeta(q)$ is a non-linear (namely parabolic)
concave function that only depends on a single positive parameter $\lambda^2$ (which quantifies the level of non linearity of $\zeta(q)$) and such that $\zeta(1) = 1$. Let us point out that the parameter $\lambda^2$ is generally referred to as the {\em intermittency coefficient} since it governs the degree of multifractality of the model, i.e., the range of the H\"older exponents that characterise the paths $X(t)$. It consequently controls the degree of appearance of volatility bursts. When $\lambda = 0$, the model is said to be {\em monofractal}, $X$ then simply corresponds to a Brownian motion which is almost everywhere of H\"older regularity $H=1/2$.

The MRW model involves a log-normal stochastic volatility, that is a multifractal random measure (MRM) $\tM_{T}$, obtained as the weak limit 
\begin{equation}
\label{MRM}
\tM_{\ell,T}(\d t) \xrightarrow[\ell \rightarrow 0]{w} \tM_{T}(\d t),
\end{equation}
where $\tM_{\ell,T}(\d t)$ is defined by 
$$
  \tM_{\ell,T}(\d t) = e^{\omega_{\ell,T}(t)} \d t,
$$
 where $\xrightarrow{w}$ stands for the weak convergence and 
the process $\omega_{\ell,T}(t)$ is Gaussian and stationary with a logarithmic covariance vanishing for lags greater than $T$ (see Eq. \eqref{covMRM}). 
Let us point out that multifractality of the limit process is obtained in the Gaussian multiplicative chaos context \cite{RV1} which implies that, at the same time  $\ell$ goes to 0, the mean (resp. variance) of $\omega_{\ell,T}$ has to go to $-\infty$ (resp. $+\infty$). Thus though the stochastic measure 
$e^{\omega_{\ell,T}(t)} \d t$ has a weak limit, the Gaussian process $\omega_{\ell,T}(t)$ does not have a limit.
We refer the reader to the beginning of Appendix \ref{App_MRM} for detailed construction of the log-normal MRM.

Since such a logarithmic decreasing covariance can be interpreted using random multiplicative cascades as the limit case where the scale ratio goes to $1$, one often refers to such a model as ``continuous cascade" \cite{bacry_kozhemyak_muzy_2013} models.
In \cite{Muzy_2002,bacry_muzy_2003}, MRM measures have been extended from log-normal statistics to any log-infinitely divisible law so that they obey the exact scaling law:
\begin{equation}{}
  \label{scaling_MRWSV}
  \E[|\delta_\tau \tM_T(t)|^q] = \E[|\tM_T(t+\tau) -\tM_T(t)|^q]\sim C_q \tau^{\zeta(q)} ,
\end{equation}
where $\zeta_q$ is the cumulant generating index of the infinitely divisible law (let us point out that it is parabolic only in the Gaussian case).

The MRM process has been used in various works since 2001 for volatility modeling. Not only it has stationary increments but it reproduces most stylized facts of volatility (including scale invariance and self-similarity properties). Moreover, it also benefits from a concise geometric construction, which allows one to easily obtain the auto-covariance function in the desired form.

\paragraph{The original RFSV model.}
In 2018, Gatheral et al. \cite{vol_is_rough} introduced a new (but related) class of models called ``rough'' fractional stochastic volatility (RFSV) models. Instead of focusing on the scaling properties of price increments, Gatheral et al. examined the regularity properties 
of the log-volatility and observed (as the case for a multifractal model) that
volatility appears to be far less regular than a Brownian motion. RFSV model quickly became a popular model. Within the RSFV framework, the volatility measure $V_{H,T}([t,t+\tau])$ of some given interval $[t,t+\tau]$ is supposed to be provided by a density measure $v(t)$ corresponding to a log-normal stationary process:

\begin{equation}\label{eq_def_vol_density_process}
V_{H,T}([t,t+\tau]) =  \int_t^{t+\tau}\! \! \!  v(s) \d s  = \int_t^{t+\tau}\! \! \!  e^{o_{H,T}(s)} \d s ,
\end{equation}
where $o_{H,T}(t)$ is a fractional Ornstein-Uhlenbeck (fOU) process that satisfies, for some $0<H<1/2$, the equation
\begin{equation}\label{eq_def_RFSV}
  \d o_{H,T}(t) = \nu \d B^H_t - \alpha (o_H(t) -m) \d t, 
\end{equation}
where $B^H_t$ is a fractional Brownian motion with Hurst parameter $H$. The parameter $\nu^2$ (resp. $m$) is the variance (resp. mean) of $o_{H,T}(t)$ and $\alpha = \frac{1}{T}$, where $T$ represents a characteristic correlation time that accounts for the typical mean reversion length of the process. Indeed, Gatheral et al. show that, for $\tau >0 $ small enough, the covariance function of $o_{H,T}(t)$ can be approximated as:
\begin{equation}
    \label{fOU_cov}
    \Cov [o_{H,T}(t), o_{H,T}(t+\tau)] \simeq \frac{\nu^2}{2} \Big( T^{2H} \Gamma(2H+1) - \tau^{2H} \Big) ,
\end{equation}
where $\Gamma$ represents the Gamma function. In \cite{vol_is_rough}, it is also shown that when $T \to \infty$, $o_{H,T}(t)$ behaves locally as  a fractional Brownian motion $B^H_t$ in the sense that, $\forall t_0$:
\begin{equation}\label{RFSV_fOU_by_fBM}
  \E [\sup_{t \in [0,t_0]} |o_{H,T}(t) - o_{H,T}(0) - \nu B^H_t|] \to 0 .
\end{equation}
This result can be of practical importance for application in finance since empirically it appears that $T$ is very large and consequently $\nu B^H_t$ can be used as a volatility model instead of the associated fOU process $o_{H,T}(t)$ as long as $t \ll T$.

Let us point out that, since the original work \cite{vol_is_rough}, many other versions of RFSV models have been introduced in the literature, each of them serving some specific purposes (making some explicit computations or estimations simpler) while keeping the main feature of the original RFSV model, i.e., the "roughness" of the volatility modelled using a fBM-like process. 
In the next section, we will introduce a new version that will enable us to unify in the same framework an RFSV model and the MRM framework.

\subsection{The log S-fBM random measure : a common framework for RFSV and MRM models}
\label{sec:def_omega}
In this section, we build the main model of this paper. This model allows us to define a common framework for RFSV and MRM models.
It is built in three steps. First we introduce a stationary version of a fractional Brownian motion, namely the S-fBM process $\{\omega_{H,T}(t)\}_t$ for $H > 0$. Then using this S-fBM process, we define the log S-fBM stochastic measure $M_{H,T}$ ($H>0$) which can be seen as a new version of a rough volatility model (RFSV). Finally, we prove that this process converges when $H$ goes to $0$ to a measure that we will refer to as $M_{0,T}$, which is shown to be an MRM.

\vskip .5cm
\noindent
{\bf Step 1/3 : Defining the S-fBM process $\{\omega_{H,T}(t)\}_t$ for $H>0$} \\
The S-fBM process $\{\omega_{H,T}(t)\}_t$ is a stationary Gaussian process and can thus be defined by its mean and its covariance function. In Appendix \ref{Appendix_geo_construction}, we provide the details of its construction by following the one proposed by Bacry \& Muzy (\cite{Muzy_2002,bacry_muzy_2003}) for building log-infinitely divisible Multifractal Random Measures (MRM). Let us point out that such a construction can also be related to the original approach proposed by Takenaka to build correlated fields (see \cite{Taqqu} and Appendix \ref{Append_proof_prop1}). 
Thus, following the construction detailed in Appendix \ref{Appendix_geo_construction}, the S-fBM process $\{\omega_{H,T}(t)\}_t$ is defined for $H>0$ as a stationary Gaussian process whose covariance function is:

\begin{equation}
\label{S-fBM_cov}
C_\omega(\tau) = \Cov[\omega_{H,T}(t), \omega_{H,T}(t+\tau)] =
\begin{cases}
\frac{\nu^2}{2}[T^{2H} - \tau^{2H} ], \; \; & \mbox{when} |\tau| <T\\
0 , \; \; & \mbox{when} |\tau| \geq T
\end{cases}
\end{equation}

\noindent
The parameter $H$  is analog to the Hurst parameter of the fBM process since it controls the ``roughness'' of the model. The variance parameter $\nu$ controls the average amplitude of the process
 and the constant $T$ is a large time scale that corresponds to the correlation scale. 
Let us point out  that
the approximated covariance provided by Eq. \eqref{fOU_cov} of the fOU process $o_{H,T}$, involved in the construction of the RFSV model, holds exactly for the S-fBM $\omega_{H,T}$ (up to a rescaling of $T$) for lags smaller than $T$. 
Both the S-fBM process and the fOU could then be regarded as stationary versions of a fBM process but, unlike the fOU process, the correlation function of S-fBM exactly vanishes for lags greater than $T$, i.e., the S-fBM values at different timestamps are independent when the distance between timestamps is large enough (i.e., greater than $T$). 

It is noteworthy that, when $T \to \infty$, one recovers the original Takenaka
construction of the fBM \cite{Taqqu} by proving that $\omega_{H,T}(t) -\omega_{H,T}(0) \to B_H(t )$.
More precisely, in Appendix \ref{Append_proof_prop1} we show that, when $T \to \infty$,  
the analog of Eq. \eqref{RFSV_fOU_by_fBM} holds for $\omega_{H,T}$:

\begin{proposition}
\label{prop_conv_to_fbm}
There exists $B_H(t)$ a fractional Brownian motion of Hurst index $H$ and unit variance at $t=1$ such that, 
$\forall t_0 > 0$, one has:
\begin{equation}
    \lim_{T \to \infty} \E [\sup_{t \in [0,t_0]} |\omega_{H,T}(t) - \omega_{H,T}(0) -  \nu B_H(t)|] = 0 \; .
\end{equation}
\end{proposition}

A direct result from the similarity in auto-covariance function is that S-fBM has the same scaling property as RFSV. According to Appendix \ref{Appendix_geo_construction}, for $\frac{\tau}{2} < T$,

\begin{equation}
  \delta_\tau\omega_{H,T}(t) = \omega_{H,T}(t+\tau)-\omega_{H,T}(t) \sim \mathcal{N}(0,\nu^2\tau^{2H}) .
\end{equation}

\noindent
It leads to the following scaling property of generalized moments, $\forall q > 0$:

\begin{equation}\label{scaling_S-fBM}
  \E[|\delta_\tau \omega_{H,T}(t)|^q] = \nu^q \frac{2^{q/2}\Gamma(\frac{q+1}{2})}{\sqrt{\pi}} \tau^{qH}\\
  =C_q \tau^{qH} \; .
\end{equation}
This means that $\log(\E[|\delta_\tau \omega_{H,T}(t)|^q])$ is linear against $\log(\tau)$ with slope $qH$.
From Kolmogorov continuity theorem it results that the paths of $\omega_{H,T}$ are continuous functions. More precisely, $\omega_{H,T}(t)$ is $\alpha-$H\"older continuous for all regularity exponents $\alpha < H$.
We especially point out that the calibration of the Hurst parameter $H$ of the log-volatility process in \cite{vol_is_rough} is based on the obtained scaling behavior \eqref{scaling_S-fBM}.

\vskip .5cm
\noindent
{\bf Step 2/3 : Defining the log S-fBM stochastic measure  $\{M_{H,T}(t)\}_t$ for $H>0$} \\
The log S-fBM stochastic measure is then defined as:
\begin{equation}
  \label{def:M}
  M_{H,T}(\d t) = e^{\omega_{H,T}(t)} \d t .
\end{equation}
Then for any interval $I$, one has:
\begin{equation}
  \label{def_lsfbm}
M_{H,T}(I) = \int_I e^{\omega_{H,T}(t)} \d t \;.
\end{equation}
Under this setting, we retrieve the so-called stationarity of volatility process, i.e.:
\begin{equation}
  \label{def:sigma2}
 \E [M_{H,T}(I)] = \sigma^2 |I|
\end{equation}
with
$$\sigma^2 = e^{m + \frac{\nu^2}{2}},$$
where $m$ (resp. $\nu^2$) is the mean (resp. variance) of $\omega_{H,T}$. The quantity $\sigma^2$ can be regarded as the variance of the price fluctuations on a unit-time interval.

\vskip .5cm
\noindent
{\bf Step 3/3 : Convergence of  $\{M_{H,T}(t)\}_t$ towards an MRM when $H$ goes to 0} \\
% The Multifractal Random Measure (MRM) \cite{MDB00,BDM01} is defined as a 
%  (weak) limit of $\ell \to 0$ of $e^{\omega_{\ell,T}}(t) \d t$ where (see Appendix \ref{App_MRM}).
As shown in Appendix \ref{App_MRM}, the MRM measure can be recovered from the log S-fBM by taking 
the limit $H \to 0$. More precisely, the following proposition holds true:
\begin{proposition}
  \label{weak-conv-prop}
  Let $M_{H,T}(t) = M_{H,T}([0,t])$ be the log S-fbm process defined by \eqref{def_lsfbm} and define
the intermittency coefficient, 
\begin{equation}
\label{deflambda}
  {\lambda^2} = {H(1-2H)}\nu^2 \; .
\end{equation}
Considering both $\lambda^2$  and the variance of the price fluctuations
$$\sigma^2 = e^{m + \nu^2/2}$$
are fixed, then, when $H \to 0$ (and consequently, $\nu^2 \rightarrow +\infty$ and $m \rightarrow -\infty$), one has
\begin{equation}
    M_{H,T}(\d t) \xrightarrow{w} \tM_T(\d t) 
\end{equation}
where $\xrightarrow{w}$ stands for the weak convergence and $\tM_T$ is a log-normal MRM (as defined by 
\eqref{MRM})
with the intermittency
coefficient $\lambda^2$ and integral scale $T$.
\end{proposition}

The proof is provided in Appendix \ref{App_MRM}. This result indicates that the MRM can be considered as a limit case of a log S-fBM and therefore could be regarded as an "extremely rough" case. Let us remark that very much like the scale parameter $\ell$ involved in the regular construction of the MRM, the parameter $H$ in the context of Proposion \ref{weak-conv-prop} can be considered as regularization parameter. However, such a regularization through $H>0$ impacts observations at all (time) scales since it explicitly breaks scale-invariance that is not the case
of regularization through $\ell$ as defined in Appendix \ref{Appendix_geo_construction}.

\vskip .5cm
\noindent
{\bf Conclusion and notations for the remaining of the paper} \\
For the sake of simplicity, in the following, the MRM $\tM_T(\d t)$ will be referred to as $M_{0,T}(\d t)$.
Thus, we can consider that we have built a class of models $M_{H,T}(\d t)$, which correspond for $H>0$ to an RFSV model and for $H=0$ to an MRM model.

\section{Second order properties of $M_{H,T}([0,t])$ and its logarithm}
\label{sec:second_order}

In section \ref{sec:gmm_estimation}, we will consider the problem of estimating the parameters of the S-fBM, namely $H,\nu^2$ (or equivalently $\lambda^2$) and $T$ through the expression of various ``statistical moments'' of the process. Among these moments, the correlation function of $M_{H,T}$ or of $Z_{H,T} = \ln M_{H,T}$ are particularly interesting since, as emphasized below, they can be approximated by simple analytical expressions.

Let us first remark that in \cite{vol_is_rough}, Gatheral et al. proposed to estimate the roughness exponent $H$ of the RFSV model (equivalently $H>0$ in the log-SfBM model) by considering the scaling of the increments of $\omega_{H,T}$ as in Eq. \eqref{scaling_S-fBM}. 
However, since $\omega_{H,T}(t)$ cannot be directly observable, they consider as a proxy of $\E[|\delta_\tau \omega_{H,T}(t)|^q]$, the observable moments:
\begin{equation}
\label{def:momZ}
  m(q,H,\tau,\Delta) = \E \Big(  |\ln M_{H,T,\Delta}(t+\tau)-\ln M_{H,T,\Delta}(t)|^q \Big), 
\end{equation}
where 
% More precisely, if we consider that $H>0$ (i.e., RFSV case of the log-SfBM model), since $\omega_{H,T}(t)$ cannot be directly observable, they consider the moments of the increments of
% \begin{equation}
%   \label{defZ}
%   Z_{H,T,\Delta}(t) = \ln M_{H,T,\Delta}(t) 
% \end{equation}
$M_{H,T,\Delta}(t)$ is the so-called integrated variance over an interval of size $\Delta$:
\begin{equation}
  \label{def:IV}
  M_{H,T,\Delta}(t) =  \sigma^2 \int_t^{t+\Delta} \! \! \! e^{\omega_{H,T}(s)} \d s \; .
\end{equation}
Thus, the exponent $H$ is measured from the scaling behavior in $\tau$ of this proxy of $\E[|\delta_\tau \omega_{H,T}(t)|^q]$ using the Eq. \eqref{scaling_S-fBM}. 

However, as emphasized below (see Section \ref{estimrough}), the estimation of $H$ based on Eq. \eqref{scaling_S-fBM} can be highly biased. 
In order to obtain an unbiased estimation of H in the framework of an RFSV model, Ref. \cite{rough_gmm} introduces a totally different framework.  The authors provide a GMM method that is based on the correlation function of $M_{H,T,\Delta}$:
\begin{equation}
  C_M(\Delta,\tau) = \E[M_{H,T,\Delta}(t) M_{H,T,\Delta}(t+\tau)] .
  % = \E [M_{H,T,\Delta}(t) M_{H,T,\Delta}(t+\tau) ] -\E [M_{H,T,\Delta}(t)]^2
\end{equation}
More precisely, they show that under peculiar conditions, its asymptotic behavior  when $\tau \gg \Delta$ can be obtained and then a GMM formula can be derived. Within the framework of various RFSV models (namely the one involving an fBM or its Riemann-Liouville variant), the authors advocate the use of this GMM method and show that it provides reliable estimates for both the roughness parameter $H$ and the variance parameter $\nu^2$.

Following this latter path, in this work, we aim at defining a GMM method for the log-SfBM framework, that works for both $H>0$ (the RFSV case) and $H=0$  (the MRM case). We thus need to establish exact or good approximations of correlation function $C_M(\Delta,\tau)$ of $M_{H,T,\Delta}$. This is the purpose of the next section (Section \ref{sec:varcor}).

Moreover, as we will see, the process $\ln M_{H,T,\Delta}(t)$ is, in some sense, close to be a Gaussian process,  consequently it is also natural to operate the GMM not on the process  $M_{H,T,\Delta}$ itself but on its logarithm $\ln M_{H,T,\Delta}(t)$. We therefore also need to establish exact or good approximations of the correlation function  of $\ln M_{H,T,\Delta}$, which is defined by :
\begin{equation}
  \label{defGammaZ}
  C_{\ln M}(\Delta,\tau) = \Cov[\ln M_{H,T,\Delta}(t),\ln M_{H,T,\Delta}(t+\tau)] \; .
\end{equation}
This is the purpose of Section \ref{sec:lnvarcor}.

% Moreover, as we will see, the process $\ln M_{H,T,\Delta}(t)$ is, in some sense, close to be a Gaussian process,  consequently the generalized moments $m(q,H,\tau,\Delta)$ 
% can all be deduced from the variance of the increments of $\ln M_{H,T,\Delta}(t)$ that itself
% can be deduced from its correlation function:
% \begin{equation}
%   \label{defGammaZ}
%   C_{\ln M}(\Delta,\tau) = \Cov[\ln M_{H,T,\Delta}(t),\ln M_{H,T,\Delta}(t+\tau)] \; .
% \end{equation}

% In this section, we therefore propose to establish exact or good approximations of second order moments of $M_{H,T}$, $C_M(\Delta,\tau)$ (Section \ref{sec:varcor}) or of the logarithm of $M_{H,T}$,  $C_Z(\Delta,\tau)$  (see Section \ref{sec:lnvarcor}).
 
\subsection{Integrated variance correlation function}
\label{sec:varcor}
In Appendix \ref{App_Corr_Sfbm} we prove the following Proposition that gives an explicit analytic formula for $C_M(\Delta,\tau)$:

\begin{proposition}
\label{prop_CM}
For any $\tau \leq T$, one has
\begin{equation}
  \label{eq:corrM1}
    C_M(\Delta,\tau)  = K_1 \big( F(\tau+\Delta)+F(\tau-\Delta)-2 F(\tau) \Big)
% -\sigma^4 \Delta^2
\end{equation}
with
\begin{equation}
\label{eq:corrM2}
  F \left(z \right) = \left \{
  \begin{array}{ll}
    z K_2^{-\frac{1}{2H}}  \biggamma(\frac{1}{2H},K_2 z^{2H})  -   K_2^{-\frac{1}{H}} \biggamma(\frac{1}{H},K_2 z^{2H}), \; \; & \mbox{when} \; \; H > 0  \\
    \frac{z^{2-\lambda^2}}{(2-\lambda^2)(1-\lambda^2)},  \; \; & \mbox{when} \; \; H = 0 
  \end{array}
  \right .
\end{equation}
where $\biggamma(a,z)$ stands for the (lower) incomplete Gamma function,
$$
 \biggamma(a,z) = \int_0^z t^{a-1} e^{-t} \d t ,
$$
and where we have denoted
\begin{eqnarray*}
K_1 & = & \frac{\sigma^4 e^{K_2 T^{2H}}}{2H}, \; \; \mbox{if} \; \; H>0 \\
& = & \sigma^4 T^{\lambda^2},  \; \; \mbox{if} \; \; H=0 \\
K_2 & = & \frac{\nu^2}{2} = \frac{\lambda^2}{2H(1-2H)} .
\end{eqnarray*}
\end{proposition}
Let us notice that when $\tau > T$, since  $M_{H,T,\Delta}(t)$ and $M_{H,T,\Delta}(t+\tau)$ are independent, one has:
$$
C_M(\Delta,\tau)  = \E[M(\Delta,\tau)]^2 = \sigma^4 \Delta^2  \; .
$$
Moreover, using the equality:
$$
 \gamma(s,z) = s^{-1} z^s e^{-z} U(1,s+1,z) ,
$$
where $U(1,s,z)$ is the Kummer's confluent hypergeometric function, the function $F(z)$ can be simply rewritten
as:
\begin{equation}
  \label{Fz}
  F(z) = \sigma^4 z^2 e^{C_\omega(z)} \left( U(1,1+\frac{1}{2H},K_2 z^{2H}) - \frac{1}{2} U(1,1+\frac{1}{H},K_2 z^{2H}) \right),
\end{equation}
where $C_\omega(z)$ is the covariance of $\omega_{H,T}$ provided by Eq. \eqref{S-fBM_cov}.
Since, when $|b| \to \infty$,
$
 U(1,b,z) \simeq 1 + \frac{z}{b} \; ,
$
and finally, when $H \ll 1$, the following approximation for $F(z)$ holds:
\begin{equation}
  \label{Fz_approx}
  F(z) \simeq \sigma^4 z^2 e^{C_\omega(z)} \left( \frac{1}{2} + \frac{3 H}{2} z^{2H} \right).
\end{equation}

\subsection{Small $\lambda^2$ approximation of the logarithm integrated variance moments}
\label{sec:lnvarcor}
In this section, our goal is to obtain analytical expressions for the moments of $\ln M([t,t+\Delta])$ instead of $M([t,t+\Delta])$. 
% like, e.g., the covariance function in Eq. \eqref{def:momZ}. 
In \cite{bacry_kozhemyak_muzy_2013} a GMM method to estimate the parameters of the MRM $\tM_T$ has been proposed relying on the expression of such logarithmic moments that
were obtained within a {\em small intermittency}, i.e. $\lambda^2 \ll 1$, asymptotic behavior. 
In fact, it is straightforward to check that all proofs and results established in \cite{bacry_kozhemyak_muzy_2013} in the limit $\lambda^2 \to 0$ for the log-normal MRM measure $\tM_T$ remain valid for $M_{H,T,\Delta}$ for $H>0$, i.e. in the log S-fBM framework introduced in this paper. Indeed, in particular by simply checking that all conditions required for MRM also hold for the log S-fBM measure $M_{H,T,\Delta}$, a direct consequence of Proposition 13 in \cite{bacry_kozhemyak_muzy_2013} is the following result:

\begin{proposition}\label{main_approx_prop}
  Let $t_1, \ldots, t_n$ be $n$ arbitrary times. The generalized moments of the logarithm of $ \Delta^{-1} M_{H,T,\Delta}(t)$
  admit the following Taylor series expansion around $\lambda^2 = 0$:
  \begin{equation}
    \E \Big[\ln\Big(\frac{M_{H,T,\Delta}(t_1)}{\Delta} \Big) \cdots \ln \Big(\frac{M_{H,T,\Delta}(t_n)}{\Delta} \Big) \Big]
    =
    \lambda^n \Delta^{-n}
    \E \Big[\Omega_{H,T,\Delta}(t_1) \cdots \Omega_{H,T,\Delta}(t_n) \Big] + o(\lambda^n) ,
  \end{equation}
  where $\Omega_{H,T,\Delta}(t)$ is the Gaussian process defined by
  \begin{equation}
    \label{def:mag}
    \Omega_{H,T,\Delta}(t) =   \frac{1}{\lambda} \int_t^{t+\Delta} \Big(\omega_{H,T}(u)- \E (\omega_{H,T}(u)) \Big) \d u .
  \end{equation}
\end{proposition}
 Within this approximation, one can directly compute $C_{\ln M}(\Delta,\tau)$ the correlation function of $\ln M_{H,T,\Delta}(t)$ as defined in \eqref{defGammaZ}.
From the definition of $\Omega_{H,T,\Delta}$ and the expression \eqref{S-fBM_cov} 
for the covariance of $\omega_{H,T}(t)$, it results: 

\begin{proposition}\label{CZ_prop}
To the first order in $\lambda^2 \ll 1$, the covariance function of $\ln M_{H,T,\Delta}$,
reads:
\begin{eqnarray}
  C_{\ln M}(\Delta,\tau) & = &  \frac{\lambda^2}{2H(1-2H)} \Delta^{-2} \int_0^\Delta \d u \int_{\tau}^{\tau+\Delta}  \Big(T^{2H} - |u-v|^{2H} \Big)  \; \; \d v \\ \label{gamma_Z}
  & = & \frac{\lambda^2}{2
    H(1-2H)} \left(T^{2H} - \frac{(\tau+\Delta)^{2H+2}+|\tau-\Delta|^{2H+2}-2 \tau^{2H+2}} {\Delta^2 (2H+1)(2H+2)}\right) + o(\lambda^2) .
  \label{eq:C_Z}
\end{eqnarray}
\end{proposition}
Let us first start with two direct consequences of these propositions
\begin{itemize}
\item When $\Delta \to 0$ one has $C_{\ln M}(\Delta,\tau) \simeq \frac{\lambda^2}{2H(1-2H)} (T^{2H}-\tau^{2H})$ which is nothing but the covariance of $\omega_{H,T}(t)$ .
\item When $H \to 0$, one recovers the expression in Proposition 10 of \cite{bacry_kozhemyak_muzy_2013} in the MRM case.
\item Proposition \ref{main_approx_prop} leads to approximating  $\ln M_{H,T,\Delta}(t)$ by a Gaussian process. 
% Within a Gaussian asumption, one can also
% obtain $C_Z$ as the logarithm of $C_M+\sigma^4 \Delta^2$. As shown by numerical evidences in the case  $\lambda^2 < 1/2$, expression \eqref{gamma_Z} turns out to provide an excellent approximation of the logarithm of expression \eqref{eq:corrM2}. Therefore considering 
% that $M_{H,T\Delta}$ is a log-normal process or the small intermittency limit  provide equivalent approximations of the covariance of $Z_\Delta$.
\end{itemize}  
 
Using this last consequence, Proposition \ref{main_approx_prop} can also be used to 
obtain an approximation to the first order in $\lambda^2$
of the moments defined in  \eqref{def:momZ}. 
Indeed, if one supposes that $\ln M_{H,T,\Delta}(t+\tau)-\ln M_{H,T,\Delta}(t)$ is a Gaussian random variable 
of variance $V(H,\tau,\Delta)$, then, 
\begin{equation}
  m(q,H,\tau,\Delta) \eql \pi^{-1/2} 2^{\frac{q}{2}} \Gamma \left(\frac{q+1}{2} \right) \Big(V(H,\tau,\Delta)\Big)^{\frac{q}{2}} \;
\end{equation} 
in which $\eql$ indicates that equality holds in the first order of $\lambda^2$.

From expression \eqref{gamma_Z}, one has when 
$\tau < T$,
\begin{eqnarray*}
  V(H,\tau,\Delta)  & = & 2 \Var(\ln M_{H,T,\Delta}(t)) - 2 C_{\ln M}(\Delta,\tau) \\
  &\eql&  \frac{\lambda^2}{H(1-2H)} \left( \frac{(\tau+\Delta)^{2H+2}+|\tau-\Delta|^{2H+2}-2 \tau^{2H+2}}{\Delta^2 (2H+1)(2H+2)}
  - \frac{2 \Delta^{2H+2}} {\Delta^2 (2H+1)(2H+2)} \right) \\
  & \eql & \lambda^2 \tau^{2H} g_H(\frac{\Delta}{\tau})
\end{eqnarray*}
with
\begin{equation}
  \label{defgh}
  g_H \left(z \right) = \frac{|1+z|^{2H+2}+|1-z|^{2H+2}-2 |z|^{2H+2}-2}{z^2 H(1-2H)(2H+1)(2H+2)}  \; .
\end{equation}

\noindent
The final expression for the moments of the increments of the measure logarithm  $\ln M_{H,T,\Delta}$ reads, in the first oder in $\lambda^2$,
\begin{equation}
  \label{mq_th}
  m(q,H,\tau,\Delta) \eql 2^{\frac{q}{2}} \pi^{-1/2} \Gamma \left(\frac{q+1}{2} \right)  \lambda^q \tau^{qH} \left[ g_H \left(\frac{\Delta}{\tau} \right)\right]^{q/2} .
\end{equation}
Let us remark that we have the following asymptotic relation:
\begin{equation}
  g_H \left(z \right)  \sim \left \{
  \begin{array}{ll}
    \frac{1}{H(1-2H)} + O(z^{2H}) \; \; & \mbox{when} \; \; z \to 0  \\
    z^{-2} \Big((1+z)^2 \ln(1+z)+(1-z)^2 \ln(1-z) -2 z^2 \ln(z)\Big) + O(H) \; \; & \mbox{when} \; \; H \to 0 
  \end{array}
  \right .
\end{equation}
and when $H>0$, one recovers that when $\Delta \to 0$ one has, to the first oder in $\lambda^2$,
\begin{equation}
  \label{mq_approx}
  m(q,H,\tau,\Delta) \eql C \tau^{qH} + O((\Delta/\tau)^{2H}), 
\end{equation}
which is the expression used to estimate $H$ in \cite{vol_is_rough} where $\ln M_{H,T,\Delta}(t) $ corresponds to the logarithm of the (daily) realized volatility.

\section{Estimation} \label{sec:gmm_estimation}

This section is devoted to the estimation of $H$ in the framework of log-SfBM. We first show (in Section \ref{estimrough}) that if $H$ is measured from the scaling behavior of $\E[|\delta_\tau \omega_{H,T}(t)|^q]$ against $\tau$ using Eq. \eqref{scaling_S-fBM} (as advocated in \cite{vol_is_rough}), the estimation of $H$ can be highly biased. 

In order to obtain an unbiased estimation of $H$, we consider two GMM based estimators in Sections \ref{sec:lh} and \ref{sec:gmm}. The first one is based on the use of the moments of the log-SfBM process itself mainly relying on the explicit covariance formula in Eq. \eqref{eq:corrM1}.
The second one is based on the use of moments of the logarithm of the log-SfBM process and involves the explicit covariance provided by Eq. \eqref{gamma_Z}.

We show that both estimators are expected to be reliable even in the "high-frequency regime" when data are only available over an interval that is smaller than the overall correlation scale $T$, i.e. in a regime when one does not expect any ergodic hypothesis to hold.

\subsection{Bias of the moment scaling method proposed in Ref. \cite{vol_is_rough}}
\label{estimrough}

In \cite{vol_is_rough}, the parameter $H$ is estimated from the scaling behavior of $\E[|\delta_\tau \omega_{H,T}(t)|^q]$ against $\tau$ as described in Eq. \eqref{scaling_S-fBM}. 
More precisely, the unobservable quantity $\E[|\delta_\tau \omega_{H,T}(t)|^q]$ is substituted by its observable proxy $m(q,H,\tau,\Delta)$ as defined in Eq. \eqref{def:momZ}, whose explicit form is worked out in Eq. \eqref{mq_th}. Then, a linear regression of $\ln m(q,H,\tau,\Delta)$ against $\ln(\tau)$ is performed in order to estimate $H$.

We can show see that this approach can lead to a significantly biased estimation of $H$. Indeed, by taking logarithm on both sides of Eq. \eqref{mq_th}, one has (using again the notation $\eql$ for equality up to the first order of $\lambda^2$)
\begin{equation}
\label{lreg}
\ln(m(q,H,\tau,\Delta))
\eql
C(q,\nu) + qH \ln(\tau) + \frac{q}{2}\ln\left(g_H(\frac{\Delta}{\tau})\right) .
\end{equation}
where the expression of $g_H(z)$ is provided in Eq. \eqref{defgh}.
Since the term $\ln(g_H(\frac{\Delta}{\tau}))$ also depends on $\ln(\tau)$, assuming, on a given range of $\tau$, that
\begin{equation}
\label{bias}
\ln(g_H(\frac{\Delta}{\tau})) \simeq B_H \ln(\tau/\Delta) + C ,
\end{equation}
the measured slope $\hat H$ in the relation $\ln m_q(H,\tau,\Delta)$ against $\ln(\tau)$ is biased as:
\begin{equation}
\hat H = H + \frac{B_H}{2},
\end{equation}
and the bias depends on both the considered range of $\tau$ and the value of $H$. 

\begin{figure}[H]
  \centering
  \includegraphics[width=0.48\textwidth]{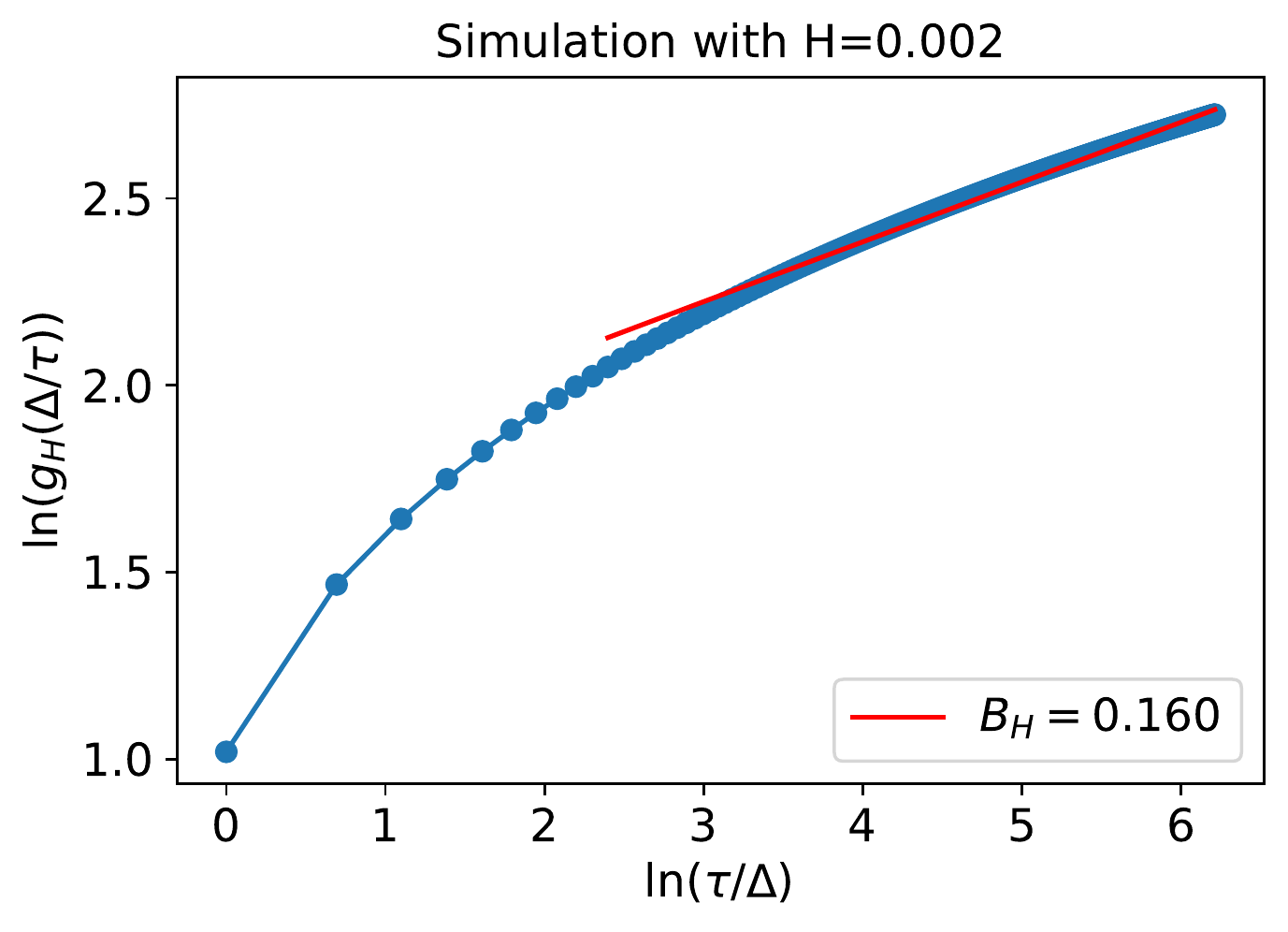}
  \caption{Estimation of the bias term $B$ (Eq. \eqref{bias}) involved in the estimation of $H$ using linear regression on Eq. \eqref{lreg}. Following Eq. \eqref{bias}, $\ln(g_H(\Delta/\tau))$ is displayed against $\ln(\tau/\Delta)$ (where $\Delta = 1$ and $\tau$ varies from 1 to 500). $B_H$ is estimated by linear regression on this curve over the range $\tau \in [10,500]$}
  \label{fig:bias}
\end{figure}

Let us illustrate this phenomenon on some numerical simulations. For that purpose, let us consider an arbitrary value $\Delta=1$ and $\tau \in [1,500]$. For the specific value $H=0.002$, Fig. \ref{fig:bias} plots $\ln(g_H(\frac{\Delta}{\tau}))$ as a function of $\ln(\frac{\tau}{\Delta}))$. We note that the behavior is, to a first approximation, assumed to be linear in the range when $\tau$ is sufficiently large in front of $\Delta$. A linear regression leads to a slope value of  $B\simeq 0.160$ which dominates the (highly biased) estimation of $H$ : $\hat H \simeq 0.08$. Using the same procedure, we checked that for different values of $H$ in the range $[0.0,0.15]$, one systematically overestimates $H$ with a bias that decreases from $0.08$ to $0.03$. 
It is noteworthy that the same kind of bias analysis has been considered by the author of Ref. \cite{vol_is_rough} themselves (see their Appendix C).

%We consider two practical scenarios. In the first scenario we consider the volatility measured in daily scale with $\Delta=8$h and $\tau \in [1,100]$ days. In the second scenario we mimic the intra-day volatility with $\Delta=5$ min and $\tau \in [5,60]$ mins. 

% For several typical $H$, the bias value, i.e $B$ is retrieved numerically. We examine the relation of $\log(g_H(z))$ against $\log(1/z)$, where $z:=\frac{h}{\tau}$. Our result suggest a significant bias in the estimated value of $H$, see Figs \ref{fig:bias} as illustration. 

%
%The so-obtained estimation value for $H$ hardly vary when $H$ varies, leading to a highly biased estimation $\hat H \simeq 0.1$ despite the true value of $H$. 
%This may explain the uniformly measured $H=0.1$ in \cite{vol_is_rough} for various assets in the Oxford-man database.
%
%\begin{table}[h]
%  \centering
%  %
%  \begin{tabular}{|c|c|c|c|c|}
%    \hline
%    \multicolumn{1}{|c|}{}         &
%    \multicolumn{1}{|c|}{$H=0.002$} & 
%    \multicolumn{1}{|c|}{$H=0.05$}   & 
%    \multicolumn{1}{|c|}{$H=0.10$}    &
%    \multicolumn{1}{|c|}{$H=0.15$}   \\ \hline
%  %
%    $B$ &  0.132 & 0.092  & 0.060 & 0.039       \\ \hline
%  %
%    $\hat H$ (estimated $H$) & 0.132  &  0.143 &  0.160  &  0.190  \\ \hline
%  \end{tabular}
%  %
%  \caption{Summary of estimated $\hat H$ against real value of $H$. {\bf ?? Moi j'enleverai la ligne avec B}}
%  \label{tbl_bias_by_scaling}
%  %
%\end{table}
%
%%In the next two sections we are going to build two GMM estimators that will prove to be hardly biased. 

%%%%%%%%%%%%%%%%%%%%%%%%%%%%%%%%%%%%%%%%%%%%%%%%%%%%%%%%%%%%%%%%%%%%%%%%%%%%%%%%%%%%

\subsection{Low versus high frequency regime for GMM estimations}
\label{sec:lh}
As already explained, our purpose is to build two GMM estimators based on the second order 
moments of the log-SfBM process $M_{H,T}$ or its logarithm. More precisely, we will consider respectively $C_M(\Delta,\tau)$ 
the correlation function of $M_{H,T,\Delta}$ (using  the explicit covariance formula 
\eqref{eq:corrM1}) and $C_{\ln M}(\Delta,\tau)$, the covariance function of $\ln M_{H,T,\Delta}$ (using the 
explicit covariance formula \eqref{gamma_Z}).
If $L$ denotes the overall size of the interval where the empirical data are available at scale $\Delta$, one can measure $M_{H,T,\Delta}(k \Delta)$ (or equivalently $\ln M_{H,T,\Delta}(k \Delta)$) for $k = 1 \ldots N$ where $N = \frac{L}{\Delta}$ 
and the estimators of previous correlation functions read:
\begin{eqnarray}
  \widehat{C_M}(\Delta,k\Delta) & = & N^{-1} \sum_{j=1}^{N-k} M_{H,T,\Delta}(j \Delta) M_{H,T,\Delta}((j+k) \Delta) \\
  \label{empiricallnM}
  \widehat{C_{\ln M}}(\Delta,k\Delta) & = & N^{-1} \sum_{j=1}^{N-k} \Big(\ln M_{H,T,\Delta}(j \Delta) - \widehat{\mu}_{\Delta} \Big)\Big(\ln M_{H,T,\Delta}((j+k) \Delta) - \widehat{\mu}_{\Delta} \Big)\\
  \widehat{\mu}_{\Delta} & = & \frac{1}{N} \sum_{k=1}^{N} \ln M_{H,T,\Delta}(k \Delta)  
\end{eqnarray}

% For that purpose we need to 
% build a vector of ``moments'' relying on various values the covariance of $M_{H,T,\Delta}$
% as in \cite{rough_gmm} or of its logarithm $Z_\Delta$ as in \cite{bacry_kozhemyak_muzy_2013}. 
% Another possibility would be to build a vector of moments based on the values of the absolute moments of
% the increments $\ln M_{H,T,\Delta}(t+\tau)-\ln M_{H,T,\Delta}(t)$, for different values of the moment order $q$ and the scale $\tau$. This would be a GMM version of the approach considered in \cite{vol_is_rough}. 

In general, GMM methods rely on some ergodic hypothesis that ensures the convergence of previous
empirical means towards the expected values. As advocated in \cite{bacry_kozhemyak_muzy_2013} or in \cite{rough_gmm}, these approaches allow one to build efficient parameter estimator in the limit $N = \frac{L}{\Delta}\to \infty$, which, when $\Delta$ is kept fixed, corresponds to $L \to \infty$. When $L\gg T$ (recall that $T$ is the correlation length of $M_{H,T,\Delta}(t)$), this ergodicity assumption can be proven to hold. We refer to such a situation as the ``low-frequency regime''.

However, as first remarked in \cite{bacry_kozhemyak_muzy_2013}, one can alternatively consider the asymptotic regime $N \to \infty$ when $\Delta \to 0$, while $L = \mathcal{O}(T)$ is fixed. This is the ``high-frequency regime''. Thus, whereas the low-frequency regime corresponds to $\Delta < T \ll L$, the second one corresponds to $\Delta \ll L = \mathcal{O}(T)$.

Let us point out that, 
as emphasized in \cite{bacry_kozhemyak_muzy_2013} and motivated by the empirical results reported in \cite{muzy_baile_bacry_2013} (see also Sec. \ref{sec:emp} below), in many practical situations and notably for financial time series, the high-frequency regime appears to fit more precisely the empirical
conditions. Notably, it appears that the correlation scale $T$ of the realized volatility always seems to be larger than the observation size $L$. For instance, in Fig. 6(b) of \cite{muzy_baile_bacry_2013},  the authors plotted the logarithm of Dow-Jones realized daily volatility from 1928 to 2011 and observed deviations far from the ``mean value'' 
that are lasting for decades. The same kind of observation can be done in Fig. \ref{fig:empRes1}(a) below.  In \cite{muzy_baile_bacry_2013}, it is also observed that the estimated correlation scale increases linearly with the observation size $L$ from a few days to several years in agreement with the hypothesis that the true correlation scale is extremely large. In such a situation, assuming that the low-frequency regime $L\gg T$ is reachable and consequently that the ergodic hypothesis holds, is clearly unrealistic.

These remarks call for developing GMM estimations in the high-frequency regime $\Delta \ll L \leq T$. 
Let us first start by noticing that from the expression of the covariance of $\omega_{H,T}$ (Eq. \eqref{S-fBM_cov}), for $t \in I$ (where $I$ is any interval such that $|I|<L$), one has 
\begin{equation}
  \{ \omega_{H,T}(t) \}_{t \in I} \stackrel{\mathcal{L}}{=} \{ \Omega + \omega_{H,L}(t)
 \}_{t \in I}  
\end{equation}
where $\stackrel{\mathcal{L}}{=}$ means an equality of all finite dimensional distributions and 
$\Omega$ is a Gaussian random variable independent of $\omega_{H,L}$ and of variance $\frac{\nu^2}{2}(T^{2H}-L^{2H})$. It thus results that we have, in any interval $I$ of size $|I|<L$,
\begin{eqnarray}
  \label{eq:ss1}
  \{ M_{H,T,\Delta} \}_{t \in I}& \stackrel{\mathcal{L}}{=} & \{ e^{\Omega} M_{H,L,\Delta}(t)\}_{t \in I} \; \; \mbox{and} \\
  \label{eq:ss2}
  \{ \ln M_{H,T,\Delta}\}_{t \in I} & \stackrel{\mathcal{L}}{=} & \{ \Omega + \ln M_{H,L,\Delta}(t)  \}_{t \in I}\; .
\end{eqnarray}
As already discussed in \cite{bacry_kozhemyak_muzy_2013} 
for the case of the MRM measure $M_{H=0,L,\Delta}$, 
these properties show that one cannot measure the parameters $T$ and $\sigma^2$ 
over an interval of size $L < T$ since by 
redefining $\sigma^2$ as $\sigma^2 e^\Omega$,  
one can always assume that $T = L$. It can also be seen on expression \eqref{eq:corrM2} that, when $\tau < T$, the large correlation scale $T$ can be absorbed in a redefinition of the variance parameter $\sigma^2$. 

\begin{center}
  \begin{figure}[h]
    \hspace*{0cm}
    \includegraphics[width=1\textwidth]{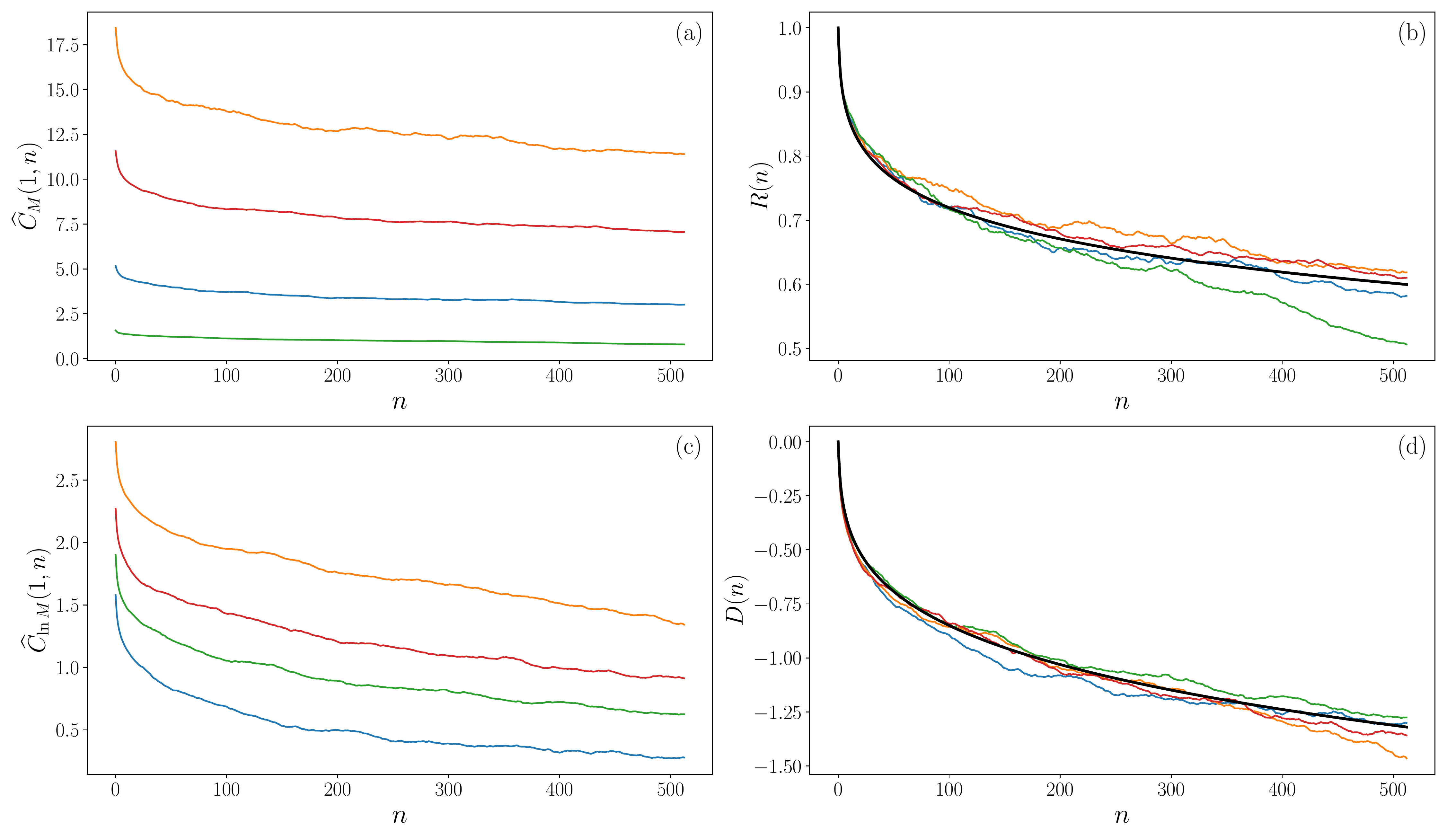}
    \caption{Estimation of the correlation functions of $M_{H,T,\Delta}$ and $\ln M_{H,T,\Delta}$ over an interval of size $L$ such that $\Delta=1 \ll L < T$. (a) $\widehat{C_M}(\Delta=1,n)$ as estimated from 4 independent realisations of $M_{H=0.1,T=2^{17},\Delta = 1}$ ($\lambda^2=0.03$) over an interval of size $L = 2^{14}$. Each estimation appears to be multiplied by an arbitrary
    random factor. (b) the ratio $R(n) = \frac{\widehat{C}_M(1,n)}{\widehat{C}_M(1,0)}$ vs. $n$. With such  normalisation,  all curves are superimposed and appear to be well fitted by the analytical expression $\widetilde{R}(n)$  represented by the bold black curve. (c)  $\widehat{C}_{\ln M}(\Delta=1,n)$ as estimated from 4 independent realisations of $M_{H=0.1,T=2^{17},\Delta = 1}$ ($\lambda^2=0.08$) over an interval of size $L = 2^{14}$. Each estimation appears to be shifted by a random term. (d) the difference $D(n) =  \widehat{C}_{\ln M}(1,n)-\widehat{C}_{\ln M}(1,0)$ vs. $n$. When shifting all curves in such a way, they are superimposed and appear to be well fitted by the analytical expression $\widetilde{D}(n)$ represented by the bold black curve.}
    \label{fig:rand_corr} 
  \end{figure}
\end{center}
If one seeks to consider correlation function based GMM estimators in the high frequency regime, one thus 
needs to study the behavior of respectively the 
estimators, $\widehat{C}_{\ln M}(\Delta,n\Delta)$, $\widehat{C}_{M}(\Delta,n\Delta)$ in the limit $\Delta \to 0$.  A rigorous study of this problem
is beyond the scope of the present paper, but we can refer to Theorem 10 of \cite{bacry_kozhemyak_muzy_2013} where the 
authors proved that, in the multifractal case ($H=0$), the behavior of $\widehat{C}_{\ln M}(\Delta,n\Delta)$ can be used to build an asymptotically unbiased and consistent estimator of $\lambda^2$ in the high frequency regime. In the present paper, we just give a sketch of proof that one can build moments functions
with vanishing fluctuations in the limit $\Delta \to 0$. 

First, let us notice that, without loss of generality, one can always perform an overall change of scale, 
$\Delta \rightarrow 1$, $L \rightarrow \frac{L}{\Delta}$, $T \rightarrow \frac{T}{\Delta}$. This amounts to
assume that $\Delta = 1$ while the limit $\Delta \to 0$ becomes  
$L,T \to \infty$ and $L = C T$ with $C  = \mathcal{O}(1)$.

Then, Appendix \ref{app:hf-regime} provides an heuristic proof of the following result:
\begin{proposition}
\label{prop_highfreq}
Suppose that $H<\frac{1}{2}$. Then, for any $n_{\max} < \infty$, $C \leq 1$, when $L = CT \to \infty$,
then, to the first order in $\lambda^2$, 
one has for all $n \leq n_{\max}$:
\begin{equation}
  \label{conv1}
  D(n) \stackrel{\mbox{def}}{=}  \widehat{C}_{\ln M}(1,n)-\widehat{C}_{\ln M}(1,0) \stackrel{P}{\longrightarrow}  \widetilde{D}_{\ln M}(n) 
\end{equation} 
where $\stackrel{P}{\longrightarrow}$ means that the convergence holds in ``in probability'' and where
\begin{equation}
  \label{expr1}
  \widetilde{D}_{\ln M}(n)    =     - \nu^2 \frac{|n+1|^{2H+2}+|n-1|^{2H+2}-2 n^{2H+2} }{2(2H+1)(2H+2)} .
\end{equation}
\end{proposition} 
Moreover, numerical experiments (see Fig.  \ref{fig:rand_corr} below) also suggest that an equivalent result holds for $\widehat{C}_{M}$, i.e.,
\begin{equation} 
\label{conv2}  
R(n) {=}  \frac{\widehat{C}_M (1,n)}{\widehat{C}_M (1,0)}\stackrel{P}{\longrightarrow}   \widetilde{R}_M (n)  
\end{equation}
with
\begin{equation}
  \label{expr2}  
  \widetilde{R}_M (n)  =   F(n+1)+F(n-1)-2 F(n)   
\end{equation}
where $F(z)$ is defined in Eq. \eqref{eq:corrM2}.

\begin{center}
  \begin{figure}[h]
    \hspace*{1.8cm}
    \includegraphics[width=0.7\textwidth]{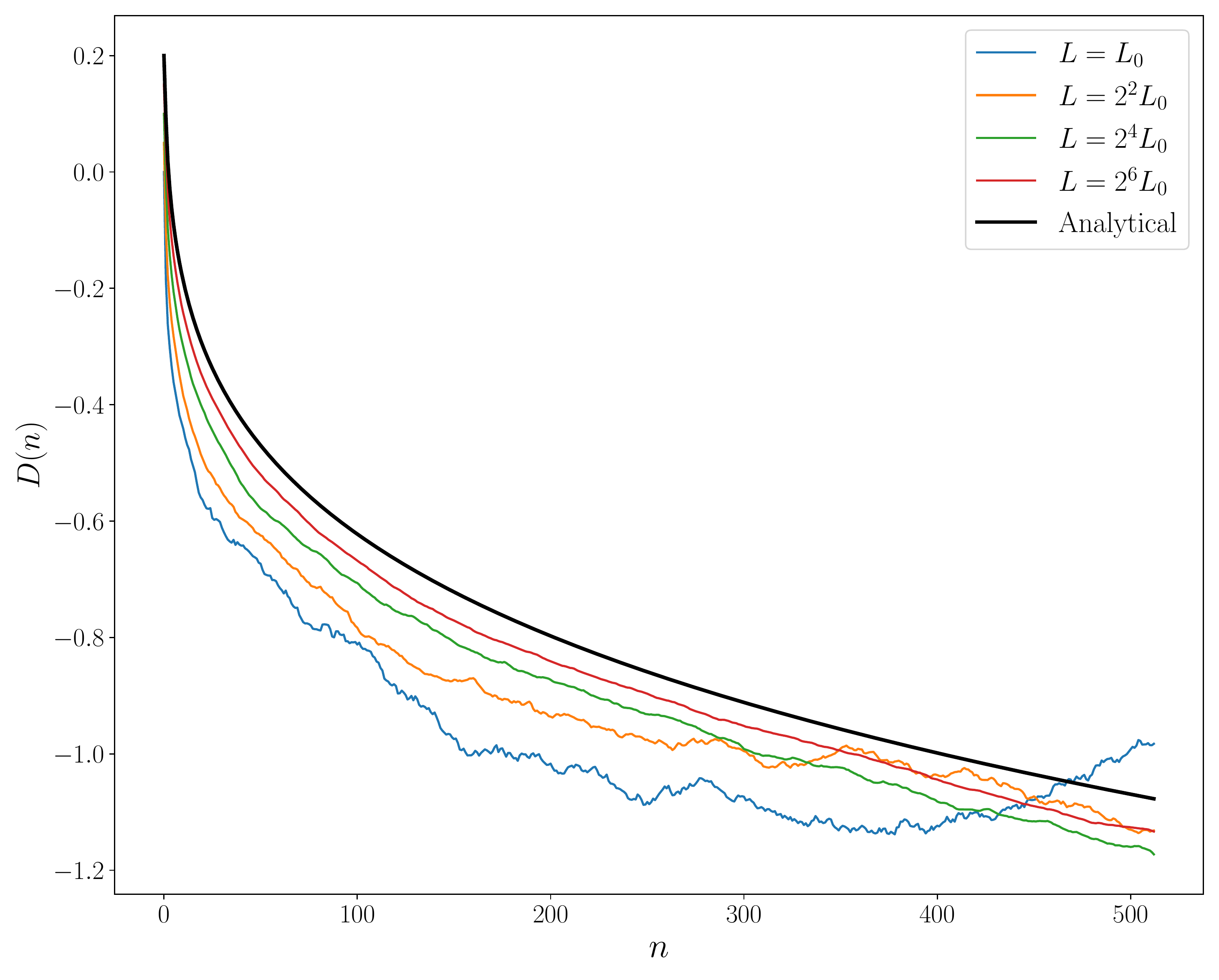}
    \caption{Estimation of the correlation function difference $D(n)$ as defined in Proposition \ref{prop_highfreq}. $D(n)$ from observations over intervals of increasing size 
    $L = L_0$, $L=4 L_0$, $L=16 L_0$ and $L = 64 L_0$ with $L_0 = 2^{12}$. The remaining parameters are $\Delta = 1$, $H=0.1$, $\lambda^2 = 0.08$ and $T = 2 L$.  One sees that as $L \to \infty$, the empirical fluctuations become smaller and smaller and the empirical estimations appear to converge towards to theoretical expression \eqref{expr1} (black bold curve). The curves have been shifted by an increasing constant for the sake of clarity.}
    \label{fig:rand_corr2}  
  \end{figure}
\end{center}

The consequence of Eqs. \eqref{conv1}, \eqref{conv2} is that, for $L$ large enough, there exist two positive random variables $K_1$ and $K_2$ such that, in the first order of $\lambda^2$):
\begin{eqnarray}
  \label{expr1_bis}
       \widehat{C}_{\ln M}(1,n) & \simeq & K_1 +  \widetilde{D}_{\ln M}(n),  \\
  \label{expr2_bis}  
   \widehat{C}_M (1,n) & \simeq  & K_2 \widetilde{R}_M (n)  .
\end{eqnarray}
Numerical illustrations of these relations are given in Figs. \ref{fig:rand_corr} and \ref{fig:rand_corr2}. 

In Fig. \ref{fig:rand_corr},
we have displayed the estimated correlation functions $\widehat{C_M}(\Delta,n \Delta)$ and $\widehat{C}_{\ln M}(\Delta,n\Delta)$ for 2 sets of 4 realisations of $M_{H,T,\Delta}(t)$ over an interval of size $L = 2^{14}$
with $\Delta = 1$, $H=0.1$, $T = 2^{17}$ and $\lambda^2=0.03$ (for $\widehat{C_M}$) or $\lambda^2=0.08$
(for $\widehat{C}_{\ln M}$). One clearly sees in Fig. \ref{fig:rand_corr}(a)
that each estimate $\widehat{C_M}$ seems to differ from the other one by a significant geometric random factor while estimates of $\widehat{C}_{\ln M}$ appear to be randomly shifted in Fig. \ref{fig:rand_corr}(c).
In order to check these assertions, we have plotted respectively the ratios $R(n)$ and the differences $D(n)$
in Figs \ref{fig:rand_corr}(b) and \ref{fig:rand_corr}(d). As expected, all the curves appear to collapse to a single curve that is well described by 
analytical expressions obtained from respectively Eq. \eqref{expr1} and \eqref{expr2} (represented by bold curves).

The asymptotic convergence of Proposition \ref{prop_highfreq} is illustrated in Fig. \ref{fig:rand_corr2}
where we have plotted $D(n)$ as defined in Eq. \eqref{conv1} as obtained from random samples of $M_{H,T,\Delta}$
with $\Delta = 1$, $H = 0.1$, $\lambda^2 = 0.08$, $T = 2L$ and $L = 2^{12}, 2^{14},2^{16}, 2^{18}$. All the curves are shifted by an arbitrary small constant for clarity purpose. As predicted by Eq. \eqref{conv1}, one sees that, as $L$ increases, the empirical curves become less and less noisy and increasingly close to the analytical expectation \eqref{expr1} (black curve).

\subsection{Defining two GMM estimators for $H$ and $\lambda^2$}
\label{sec:gmm}
We are now ready for defining two GMM estimators for $H$ based respectively on the moments of $M_{H,T,\Delta}$ or its logarithm in the high-frequency limit. By using the previously established expressions \eqref{expr1} and \eqref{expr2} for the empirical correlation function
$\widehat{C}_{\ln M}(1,n)$ and  $\widehat{C}_{M}(1,n)$, one can devise two GMM methods along the same line as the methods proposed respectively in \cite{bacry_kozhemyak_muzy_2013} and \cite{rough_gmm}.

As explained in the previous section (Section \ref{sec:lh}), in the high-frequency regime, estimations of $T$ or $\sigma^2$ are unreachable.  Thus, hereafter, we  consider exclusively the problem of estimating the values of the parameters
$H$ and $\lambda^2$ (or alternatively $\nu^2$) using one of 
the following two sets of moments:
\begin{eqnarray*}
  \mbox{GMM}_M: \; \; {\cal M}_1 & = & \Big( \widehat{C_M}(1,j_1)-\widetilde{C}_M(1,j_1), \ldots, \widehat{C_M}(1,j_Q)-\widetilde{C}_M(1,j_Q) \Big) \;, \\
  \mbox{GMM}_{\ln M}: \; \; {\cal M}_2 &  = & \Big( \widehat{C_{\ln M}}(1,j_1)-\widetilde{C}_{\ln M}(1,j_1), \ldots, \widehat{C_{\ln M}}(1,j_Q)-\widetilde{C}_{\ln M}(1,j_Q) \Big) \;, 
\end{eqnarray*}
where $Q$ is the number of moments, $j_1,j_2,\ldots, j_Q$ different time indices, $\widehat{C_M}$ and $\widehat{C}_{\ln M}$ are the empirical estimators of respectively $C_M$ and $C_{\ln M}$ and $\widetilde{C}_{M}, \widetilde{C}_{\ln M}$ the following analytical expressions:
\begin{eqnarray}
  \label{expr1_ter}
  \widetilde{C}_{\ln M}(1,n) & = & K_1 +  \widetilde{D}_{\ln M}(n) + V_1 \delta_n  \\
  \label{expr2_ter}  
  \widetilde{C}_M (1,n) & =  & K_2 \widetilde{R}_M (n)  
\end{eqnarray}
where $K_1,K_2$ and $V_1$ are 3 random positive constants and $\delta_n$ stands for the Kronecker function.
Notice that the term $V_1 \delta_n$ allows one to account for the eventual presence of a white noise (of variance $V_1$ superimposed to $\ln M_{L,T,1}(t)$ as described in ref. \cite{rough_gmm}.

\subsection{Numerical illustrations and empirical performances of the GMM methods}
In order to verify our approach and compare the performances of $\mbox{GMM}_M$ and $\mbox{GMM}_{{\ln}M}$, we have carried out various numerical experiments.
However, since historical volatility is not directly observable in financial markets, in order to consider a more realistic scenario, we decided to run the experiments directly on a price model. 
We consider that a ``price'' $X_t$ is modelled by a Brownian motion whose variance is a  log S-fBM measure $\d M$, i.e., 
\begin{equation}
\label{eq:model_price}
  dX_t = e^{\omega_{H,T}(t)} \d B_t = \frac{M_{H,T}(\d t)}{\d t} \d B_t \; ,
\end{equation}
where $M_{H,T}$ is the log-fBM defined in \eqref{def:M} while $B_t$ is a Brownian motion
independent of $M_{H,T}$. Let us notice that, when $H = 0$, $X_t$ is precisely the MRW process 
introduced in \cite{MDB00,BDM01}. Let us also remark that, in many respects (e.g., by rewriting $\omega_{H,T}$ as a stochastic integral of the form $\int_{-\infty}^t g(t-s) dW_s$), the model \eqref{eq:model_price} can be seen as a peculiar non skewed variant a the rough Bergomi model introduced 
in \cite{Bayer_Friz_Gatheral_rBergomi}.

Alternatively, an equivalent definition of $X_t$ can be obtained using a time-warp of the Brownian motion: 
\begin{equation}
  X_t = B(M_{H,T}(t)) \; .
\end{equation}
Within this framework, $M_{H,T}(t)$ is called the (stochastic) volatility of $X_t$.
If one does not observe directly $M_{H,T}(\d t)$ but only the process $X_t$, as emphasized notably in \cite{BN02}, a proxy of the integrated volatility over an interval of size $\Delta$ is provided by an estimation of the quadratic variation of $X$:
\begin{equation}
  \hM_{H,T,\Delta}(t) = \sum_{i=1}^n \left(X_{t+\frac{i \Delta}{n}}-X_{t+\frac{(i-1)\Delta}{n}} \right)^2 \; .
\end{equation}
As shown in \cite{BN02} (see also \cite{rough_gmm}), as $n \to \infty$, under mild conditions, $\hM_{H,T,\Delta}  \to M_{H,T,\Delta}$ while even for moderate $n$, 
$\hM_{H,T,\Delta}(t)$ and $\ln \hM_{H,T,\Delta}(t)$ provide excellent approximations 
of the integrated volatility and its logarithm. For the purpose of this paper, we have checked that $n = 32$ is sufficient to disregard any significant difference between $\hM$ and $M$.

We simulated independent samples of S-fBM processes and the associated processes $X_t$ 
with $H = 0.02$, $H = 0.08$ and $H = 0.15$, with 2 different values of $\lambda^2$, namely $0.02$ and $0.1$. We chose $T = 2^{17}$, $L = 2^{14}$ and fixed arbitrary $\sigma^2 = 1$.

For all these parameters we run both $\mbox{GMM}_M$ and $\mbox{GMM}_{{\ln}M}$ estimators with $Q = 19$ 
and $\{\tau_k\}_{k=0,\ldots,18} = \lf \sqrt{2^{k}} \rf$.
Our GMM implementations closely follow the one detailed in \cite{rough_gmm} and notably the error covariance is estimated using the Newey-West HAC type estimator with a lag $L^{1/3}$ and the initialisation is performed using the scaling estimator provided in \cite{vol_is_rough}. We used the {\em L-BFGS-B} minimisation algorithm as provided by {\em scipy.optimize} library in Python but we find similar results using alternative methods.

\begin{center}
  \begin{figure}[t]
    \hspace*{-0.3cm}
    \includegraphics[width=1\textwidth]{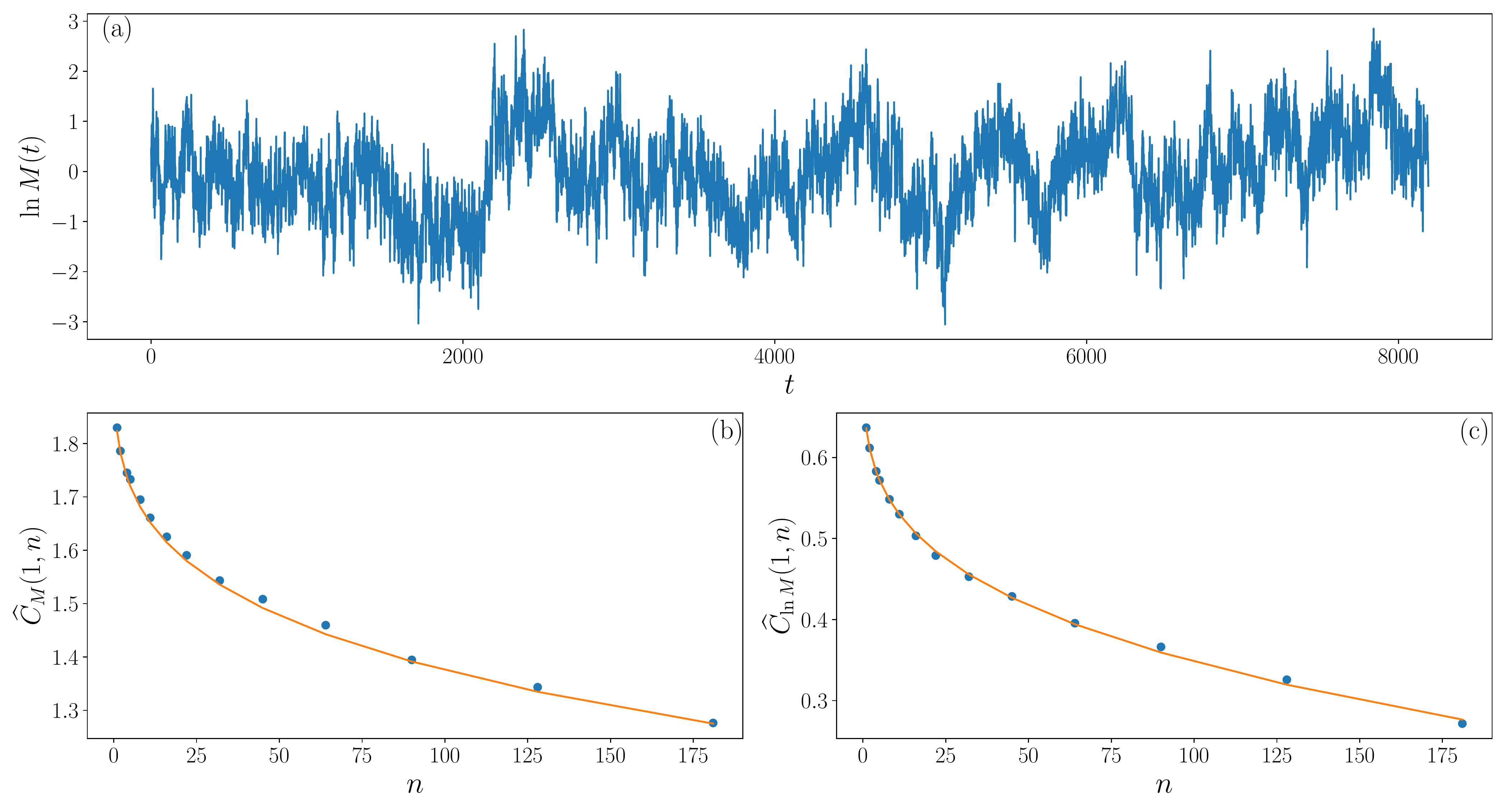}
    \caption{GMM estimations of $H$ and $\lambda^2$ (a) A sample of length $L=2^{14}$ of the ``log-volatility", $\ln \hM_{H,T,\Delta}(t)$, with $n=32$, $\Delta = 1$, $H = 0.08$, $\lambda^2 = 0.1$ and $T = 2^{17}$. (b) Best GMM fit of ${\widehat C}_M(\tau)$ as a function of $\tau$. (c) Best GMM fit of ${\widehat C}_{\ln M}(\tau)$ as a function of $\tau$.}
    \label{fig:GMM_ex}
  \end{figure}
\end{center}
\vspace*{-1cm}

In Fig. \ref{fig:GMM_ex}, are displayed a fit of respectively $\widehat{C_M}(1,n)$
and $\widehat{C}_{\ln M}(1,n)$ using expressions \eqref{expr1} and \eqref{expr2}
with the estimated GMM parameters for a sample of length $L = 16384$ with $H = 0.08$
and $\lambda^2 = 0.02$ or $\lambda^2 = 0.1$.
Our estimation results are summarised in Table \ref{tbl_1} where we reported the obtained mean values and standard deviation of estimated $H$ and $\lambda^2$ for each set of parameters. 

\begin{table}[H]
  \centering
  \begin{tabular}{|l|l|l|l|l|}
    \hline
    \multicolumn{1}{|c|}{$\lambda^2 = 0.02$}         &
    \multicolumn{1}{|c|}{$H = 0$} & \multicolumn{1}{|c|}{ $H=0.02$}        & \multicolumn{1}{|c|}{$H=0.08$}         &
    \multicolumn{1}{|c|}{$H=0.15$}   \\ \hline
    $\widehat H$ ($\mbox{GMM}_{M}$) & 0.010 (0.01) & 0.007 (0.015) & 0.077 (0.033) & 0.146 (0.05) \\ \hline
    $\widehat H$ ($\mbox{GMM}_{{\ln}M}$) & 0.010 (0.01) & 0.018 (0.015) & 0.082 (0.02) & 0.153 (0.02) \\ \hline 
    $\widehat \lambda^2$ ($\mbox{GMM}_{M}$) &  0.010 (0.01) & 0.010 (0.01) & 0.018 (0.006) & 0.021 (0.005)       \\ \hline
    $\widehat \lambda^2$ ($\mbox{GMM}_{{\ln}M}$) & 0.019 (0.001) &  0.020 (0.001)     &  0.019 (0.002)       &  0.020 (0.002)     \\ \hline \hline
    \multicolumn{1}{|c|}{$\lambda^2 = 0.1$} & 
    \multicolumn{1}{|c|}{$H=0$} & \multicolumn{1}{|c|}{$H=0.02$}         & \multicolumn{1}{|c|}{$H=0.08$}         &
    \multicolumn{1}{|c|}{$H=0.15$}   \\ \hline
    $\hat H$ ($\mbox{GMM}_{M}$) &  0.010 (0.02)& 0.018 (0.02) &  0.11 (0.22) & 0.16 (0.26) \\ \hline
    $\hat H$ ($\mbox{GMM}_{{\ln}M}$) & 0.010 (0.01) & 0.02 (0.01) & 0.078 (0.02) & 0.16 (0.02)  \\ \hline
    $\widehat \lambda^2$ ($\mbox{GMM}_{M}$) & 0.08 (0.03) & 0.08 (0.02)   & 0.09 (0.045)    &   0.08 (0.07)     \\ \hline
    $\widehat \lambda^2$ ($\mbox{GMM}_{{\ln}M}$) & 0.095 (0.001) & 0.10 (0.005)   & 0.10 (0.008)    &   0.10 (0.008)  \\ \hline
  \end{tabular}
  \caption{Summary of $\mbox{GMM}_{M}$ \& $\mbox{GMM}_{{\ln}M}$ estimation performances. For each parameter set, we report the mean values and standard deviations as obtained from estimations realized on 50 independent samples of length $L = 2^{14}$ of log S-fBM stochastic volatility model.}
  \label{tbl_1}
\end{table}

\vspace*{-1cm}

\begin{center}
	\begin{figure}[h]
		\hspace*{0cm}
		\includegraphics[width=1\textwidth]{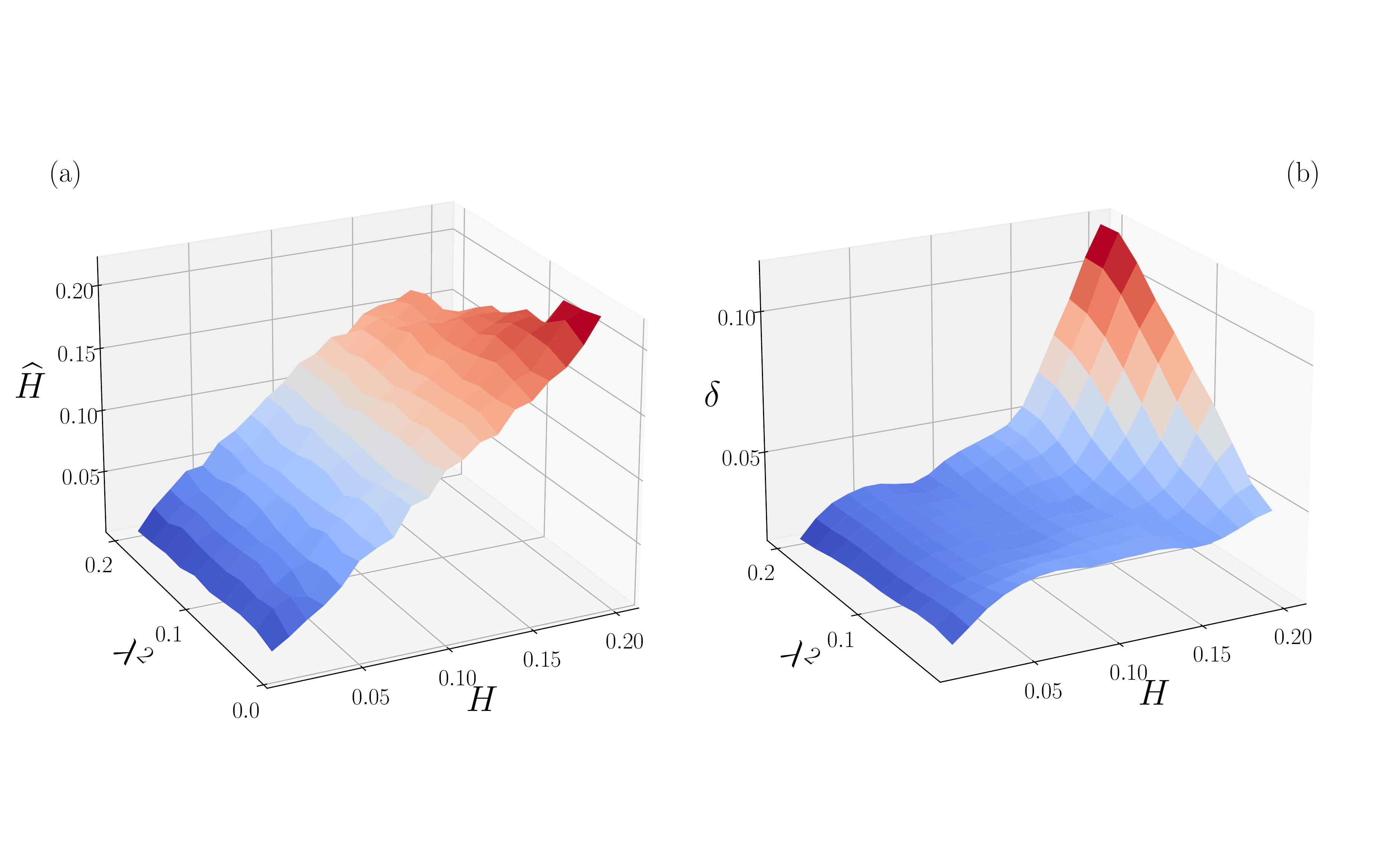}
		\vspace*{-2cm}
		\caption{Surface plot of the Hurst parameter $H$ estimation error as a function 
			of the parameters $H$ and $\lambda^2$. For each pair of parameters values, the considered log S-fBM stochastic volatility series are of length $L = 4096$, with $T=2^{17}$ and the averages $E(.)$ are computed over 50 independent draws. (a) $E(\widehat{H})$, the average $\mbox{GMM}_{{\ln}M}$ estimation $\widehat{H}$ is plotted as as function of $H$ and $\lambda^2$. On observes a significant negative bias for largest values of $H$ and $\lambda^2$ ($H \geq 0.15$ and $\lambda^2 \geq 0.1$). (b) The error
			$\delta = \sqrt{E[(H-\widehat{H})^2]}$  as a function of $H$ and $\lambda^2$. On can see that the error remains small (i.e. less than 0.03) as long as $H<0.15$ and $\lambda^2 < 0.1$.}
		\label{fig:Fig_Error_Surf}
	\end{figure}
\end{center}

\begin{center}
	\begin{figure}[h]
		\hspace*{1.5cm}
		\includegraphics[width=0.8\textwidth]{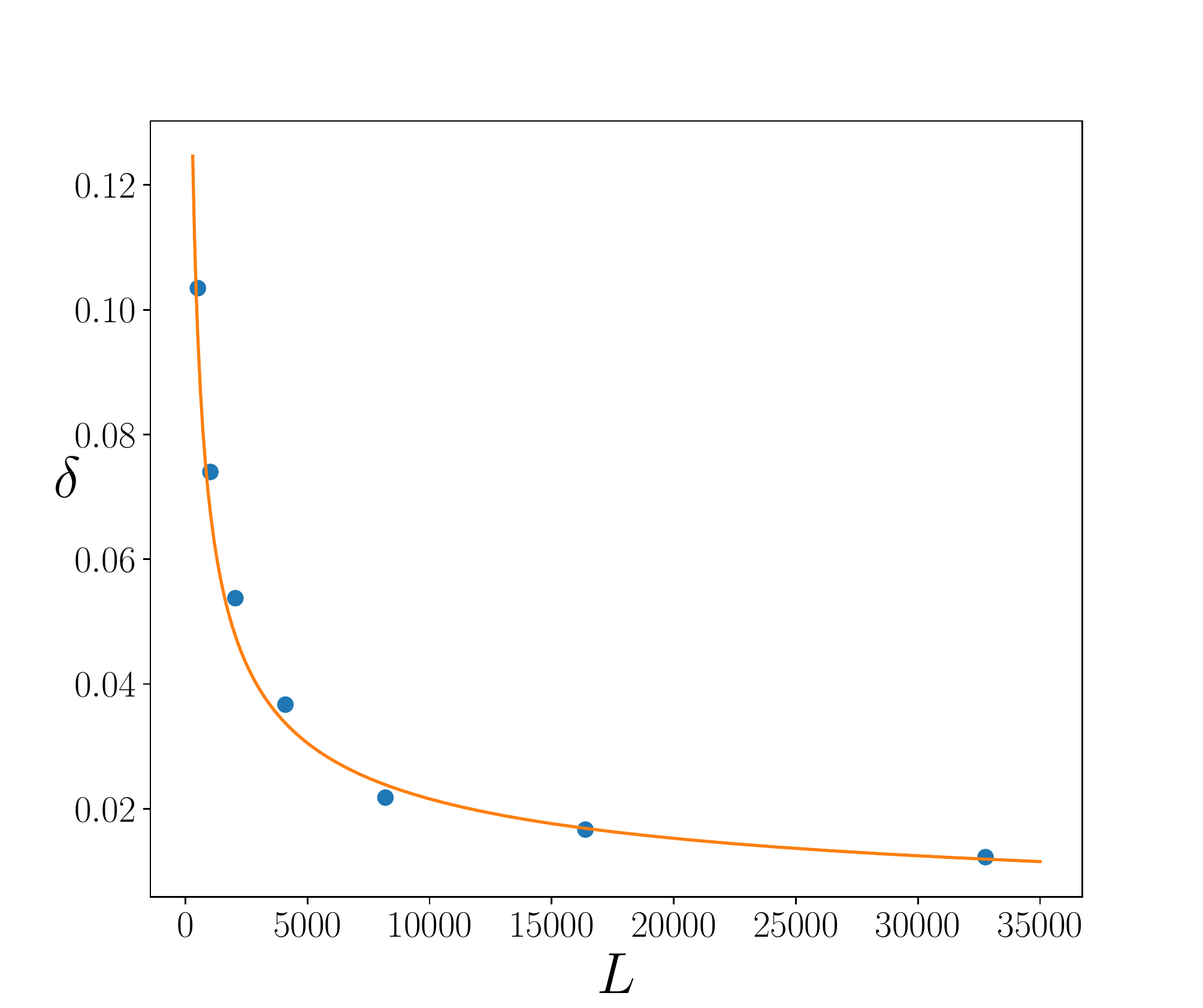}
		\vspace*{0cm}
		\caption{$\mbox{GMM}_{{\ln}M}$ Hurst exponent estimation error $\delta$ as a function of the sample size $L$. For each size $L \in [2^{9},2^{10},2^{11},2^{12},2^{13},2^{14},2^{15}]$, we have generated 200 realizations of log S-fBM stochastic volatility series of length $L$ with $H = 0.1$, $\lambda^2 = 0.03$ and $T = 4 L$. Empirical error are represented by symbols ($\bullet$) while the solid line represents the curve $\delta  = K L^{-1/2}$.}
		\label{fig:Fig_Error_vs_Size}
	\end{figure}
\end{center}

We clearly see that the
$\mbox{GMM}_{{\ln}M}$ method relies on logarithms of integrated volatilities outperforms the $\mbox{GMM}_{M}$  method built on integrated volatilities. 
This latter approach appears to have significantly 
larger bias and variance errors notably for very small $H$ values.
$\mbox{GMM}_{{\ln}M}$ method provides more reliable estimates and in particular one sees that the error on $\lambda^2$ is very small for all sets of parameters. In order to 
better illustrate the variations of $\delta$, the estimation error on $\widehat{H}$, as a function of the S-fBm parameters, we have displayed it as a surface plot in Fig. \ref{fig:Fig_Error_Surf} for a size $L = 2^{12}$ comparable to ones of the empirical time series considered in section \ref{sec:emp}. One can see that the error increases when both $H$ increases and $\lambda^2$ increase but remains rather small in the domain $0 \leq H \leq 0.15$ and $0 < \lambda^2 \leq 0.1$. As far as the estimation error of $\lambda^2$ is concerned, we have observed similar results though with a relative error smaller than $0.1$ over the whole domain.

Regarding the scaling of the error with respect to the sample size $L$, by considering, besides $L = 2^{12}$ as in Fig. \ref{fig:Fig_Error_Surf} or $L = 2^{14}$ as in Table \ref{tbl_1}), various sample lengths ($L=2^{9},2^{10},2^{11},2^{12},2^{13},2^{14},2^{15}$) with $T = 4 L$, we checked that, as predicted by Prop. \ref{prop_highfreq}, the estimation errors vanishes when $L$ increases. As illustrated in Fig. \ref{fig:Fig_Error_vs_Size}, empirically, it appears that, even in the high-frequency regime, the error behaves as $L^{-1/2}$.

\begin{center}
  \begin{figure}[h]
    \vspace*{-1cm}
    \hspace*{1.2cm}
    \includegraphics[width=0.8\textwidth]{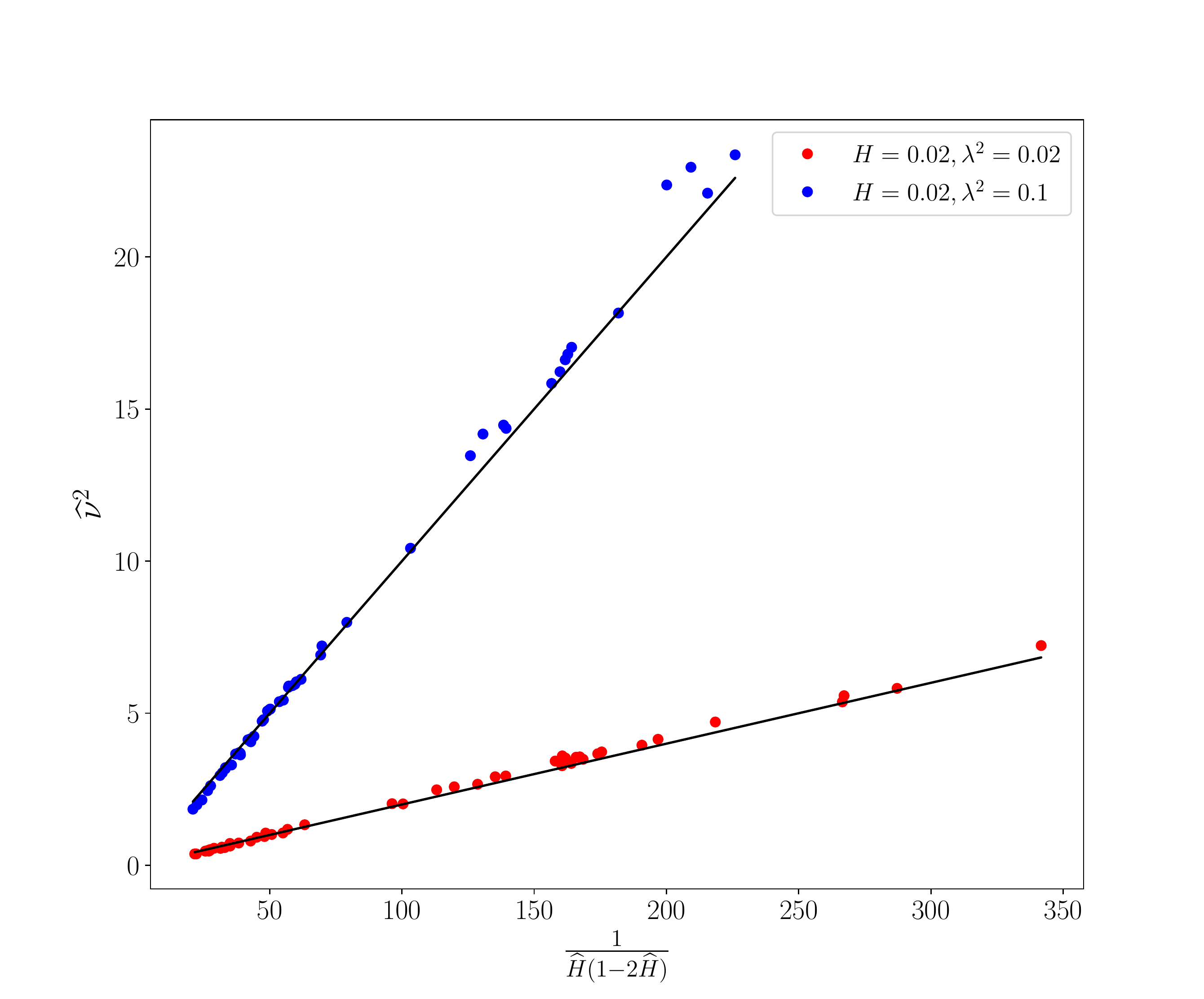}
    \caption{$\mbox{GMM}_{{\ln}M}$ estimation of $\nu^2$ as a function of $\frac{1}{{\widehat H}(1-2{\widehat H})}$. Each point corresponds to one estimation on a sample of length $L = 2^{14}$. The parameters are $H = 0.02$ and $\lambda^2 = 0.02$ (red symbols) or $\lambda^2 = 0.1$ (blue symbols). The straight lines represent the fit provided by Eq. \eqref{eq:nu_vs_H}.}
    \label{fig:nu2_estim}
  \end{figure}
\end{center}

Finally, let us emphasise that the reported estimations were obtained by
estimating $H$ and the variance parameter $\nu^2$ from which 
$\lambda^2$ is estimated using Eq. 
\eqref{deflambda}. We checked that estimating directly $\lambda^2$ instead of deriving it from $\nu^2$, provides the same results. However we observed that the errors on $\nu^2$ are much larger than the errors on $\lambda^2$. More precisely, it appears that, for a fixed $\lambda^2$, the measured bias is strongly related to $\widehat H$ as precisely expected from :
\begin{equation}
  \label{eq:nu_vs_H}
  {\widehat \nu^2} \simeq \frac{\lambda^2}{{\widehat H}(1-2{\widehat H})} .
\end{equation}
This is illustrated in Fig. \ref{fig:nu2_estim} in which two experiments where run with $H = 0.02$. For the first one we chose $\lambda^2 = 0.02$ and for the other one we chose
$\lambda^2 = 0.1$. 

\vspace*{-0.5cm}

For each sample, we have reported ${\widehat \nu^2}$ as a function of $\frac{1}{{\widehat H}(1-2{\widehat H})}$ estimated by $\mbox{GMM}_{{\ln}M}$ method.
One can easily see that in each case ($\lambda^2 = 0.02$ or $\lambda^2 = 0.1$),
one observes a very large dispersion on ${\widehat \nu^2}$ (whose expected values should be respectively $\nu^2 = 1.04$ and $\nu^2 = 5.2$) that, however, strikingly appears to be proportional 
to $\frac{1}{{\widehat H}(1-2{\widehat H})}$ (which, 
when $H$ is very small, has a large dispersion). As shown by the linear fits
predicted by Eq \eqref{eq:nu_vs_H} (continuous line in Fig. \ref{fig:nu2_estim}), the proportionality constant is precisely the value of the intermittency coefficient $\lambda^2$ for which the estimation is quite accurate. These observations suggest that while $\lambda^2$ can be estimated with a very small error, this is not at all the case of $\nu^2$, when $H \ll 1$.  The intermittency coefficient $\lambda^2$ appears to be a much more reliable quantity than the variance $\nu^2$ of the log volatility.
This can be easily explained by the fact that, in order for the S-fBM measure to converge when $H \to 0$ (towards the MRM $\tilde{M}$), one has to choose a variance proportional to $1/H$. Therefore, in the moment estimation method, in order to match the empirical covariance values when the estimated $H$ is very small, the parameter $\nu^2$ must scale as $H^{-1}$.

%%%%%%%%%%%%%%%%%%%%%%%  Application %%%%%%%%%%%%%%%

\begin{center}
	\begin{figure}[ht]
		\hspace*{0.2cm}
		\includegraphics[width=0.9\textwidth]{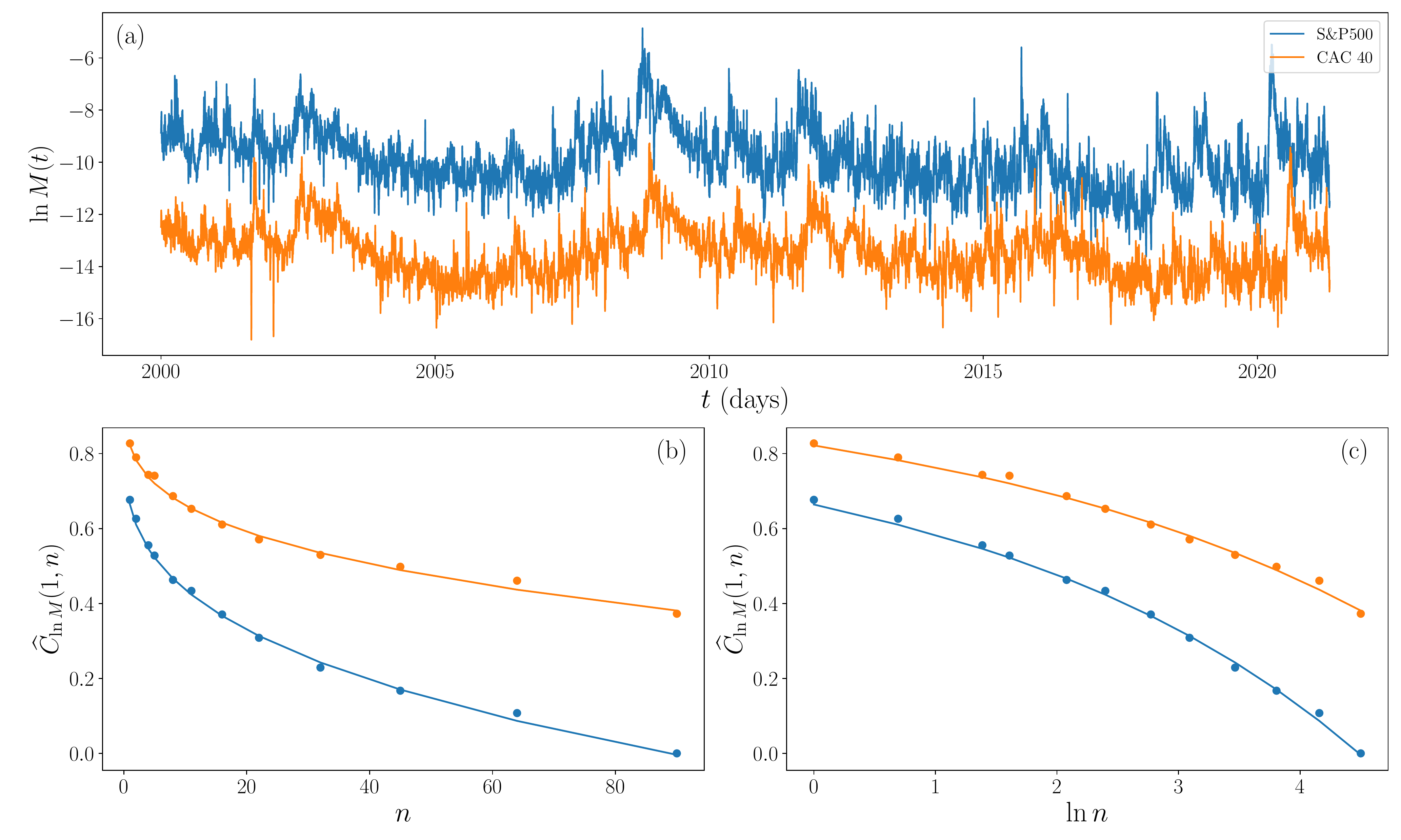}
		\caption{$\mbox{GMM}_{{\ln}M}$ estimation of daily volatility of S\&P500 and CAC40 indices from Oxford-Man dataset 
			(a) log-realized bipower-variation from January 2000 to March 2021. 
			(b) (resp. (c)) : the dots represent the estimations ${\widehat C}_{\ln M}(\tau,\Delta)$ of the corresponding correlation function ${\widehat C}_{\ln M}(\tau,\Delta)$ for each index as a function of $\tau$ (resp. $\ln \tau$). The plain lines correspond to the $\mbox{GMM}_{{\ln}M}$ fits. The so-obtained estimated values of $H$ are respectively $H \simeq 0.14$ (for S\&P) and $H \simeq 0.13$ (for CAC40). CAC40 curves in Figures (a),(b) and (c) have been arbitrary shifted for the sake of clarity.}
		\label{fig:empRes1}
	\end{figure}
\end{center}

\begin{center}
	\begin{figure}[h]
		\hspace*{0.2cm}
		\includegraphics[width=0.9\textwidth]{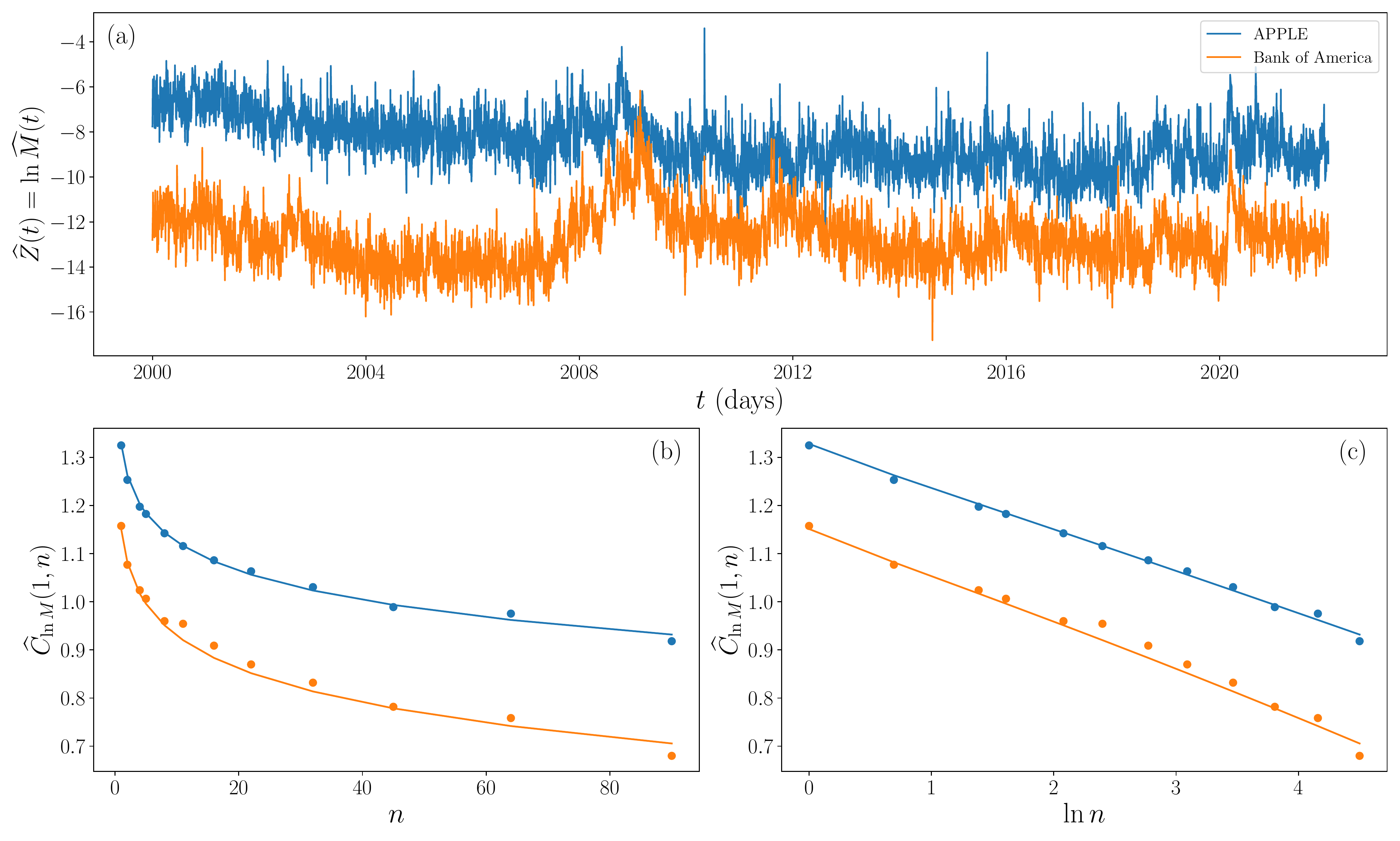}
		\caption{$\mbox{GMM}_{{\ln}M}$ estimation of daily volatility of Apple and Bank of America from Yahoo Finance dataset.
			(a) log-realized Garman-Klass estimation of volatility \cite{GK80} from January 2000 to December 2021.
			(b) (resp. (c)) : the dots represent the estimations ${\widehat C}_{\ln M}(\tau,\Delta)$ of the corresponding correlation function ${\widehat C}_{\ln M}(\Delta,\tau)$ for each index as a function of $\tau$ (resp. $\ln \tau$. The plain lines correspond to the $\mbox{GMM}_{{\ln}M}$ fits. The so-obtained estimated values of $H$ are respectively $H \simeq 0.01$ (for Apple) and $H \simeq 0.022$ (for Bank of Am.). Apple curves in Figures (a),(b) and (c) have been arbitrary shifted for the sake of clarity.}
		\label{fig:empRes2}
	\end{figure}
\end{center}

\vspace*{-1cm}
\begin{center}
	\begin{figure}[h]
		\hspace*{2cm}
		\includegraphics[width=0.7\textwidth]{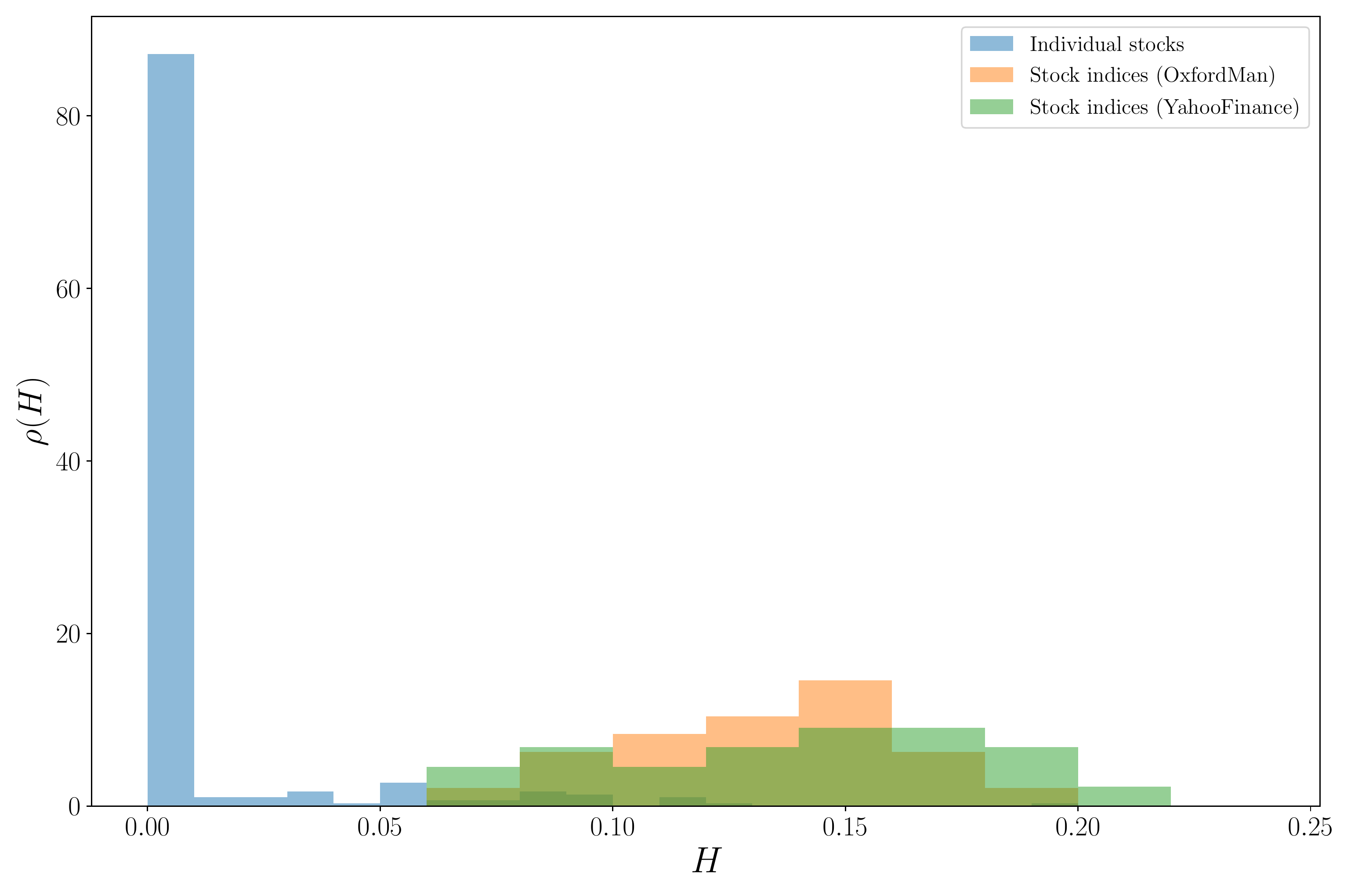}
		\vspace*{-0.5cm}
		\caption{Estimation of the probability density distribution of Hurst exponent estimation $\hat H$ for the 296 individual stocks (blue bars) of the YF database and for the 24 stock market indices  of the OM database (orange bars) or YF database (green bars).}
		\label{fig:Fig_5_Histo}
	\end{figure}
\end{center}

\vspace*{-1cm}
\section{Application to realized volatility of asset returns}
\label{sec:emp}
In this section we consider the application of the estimator of the former section to characterise the roughness exponent $H$ and the intermittency coefficient $\lambda^2$ of realized volatility associated with various assets.
Section \ref{sec:gmm_estimation} suggests that the GMM estimator $\mbox{GMM}_{{\ln}M}$ outperforms the other candidate $\mbox{GMM}_{M}$. This is why we exclusively consider the $\mbox{GMM}_{{\ln}M}$ applied to various empirical daily volatility data. Our study is based on 2 datasets containing respectively stock market indices and  individual stock prices:

\paragraph{\em Oxford-Man Institute of Quantitative Finance Realized Library (OM)\\}  
The Oxford-Man Institute's Realized Library\footnote{http://realized.oxford-man.ox.ac.uk/data}, contains historical records of various estimators of daily realized volatility of several stock indices. This dataset is widely used in various empirical studies and in particular, it was used as a benchmark database in many former studies on rough volatility (see e.g. \cite{vol_is_rough, rough_gmm}). So we apply $\mbox{GMM}_{{\ln}M}$ estimator to analyse the daily volatility time series associated with 24 major stock market indices considered in \cite{rough_gmm}. Following this latter work, in the following, we only report obtained results when using bipower variation volatility estimator but we have checked that the same results are obtained when using realized variance estimators at scale $5$ min or $10$ min. Two $\mbox{GMM}_{{\ln}M}$ estimations are illustrated in Fig. \ref{fig:empRes1} : one on CAC40 data and one on S\&P500 data. The corresponding daily historical volatilities (using bipower-variation estimator) are illustrated in Fig. \ref{fig:empRes1}(a).
We observe that over the 20-years period, the volatilities of S\&P 500 and CAC40 are strongly correlated. One can also notice that some of the correlated departures from the mean value are lasting several years. This observation seriously questions any ergodic hypothesis that would result from short-term correlations as assumed in many papers (see, e.g., \cite{vol_is_rough,rough_gmm}). 
Figs \ref{fig:empRes1}(b) (resp. (c)) displays the corresponding estimated correlation functions $\hat C_{\ln M}(\Delta,\tau)$ as a function of $\tau$ (resp. $\ln \tau$) and their $\mbox{GMM}_{{\ln}M}$ fits.
The so-obtained estimations for $H$ are $H \simeq 0.14$ (for S\&P) and $H \simeq 0.13$ (for CAC40). For both indices, we estimate $\lambda^2 \simeq 0.05$.

\paragraph{\em Yahoo Finance database (YF) \\}
We collected historical daily open, high, low, and close price time-series of 296 individual stocks and also of a set of 24 stock indices from Yahoo Finance\footnote{http://YahooFinance.com}. Stocks were taken from either the S\&P 500 index (historical data from 1985-01-01 to 2021-12-31) or the CAC 40 index ((historical data from 2000-01-01 to 2021-12-31) while the indices were chosen as being those we considered in the OM database (over the period 200-01-01 to 2021-12-31). For each asset, we constructed a proxy of the daily volatility using Garman-Klass (GK) estimator described in \cite{GK80}. This allows us, for any individual stocks or any index, to perform
a $\mbox{GMM}_{\ln M}$ estimation of $H$ and $\lambda^2$ from the estimated log-volatility time series. 
As illustrated in Figs \ref{fig:Fig_5_Histo} and  \ref{fig:Fig_5_ln} below, on stock indices our results using Oxford-Man realized volatility and GK volatility estimation from YF data provide results that are fully consistent. In Fig. \ref{fig:empRes2}, following the exact same structure as Fig. \ref{fig:empRes1}, we illustrated the estimation procedure with the examples of Apple and Bank of America realized volatility. Again, we observe that volatility fluctuations seem to be long-term correlated. For the selected two stocks, the estimated values of $H$ are respectively $0.01$ and $0.02$.

\begin{center}
	\begin{figure}[h]
		\hspace*{-0.3cm}
		\includegraphics[width=1\textwidth]{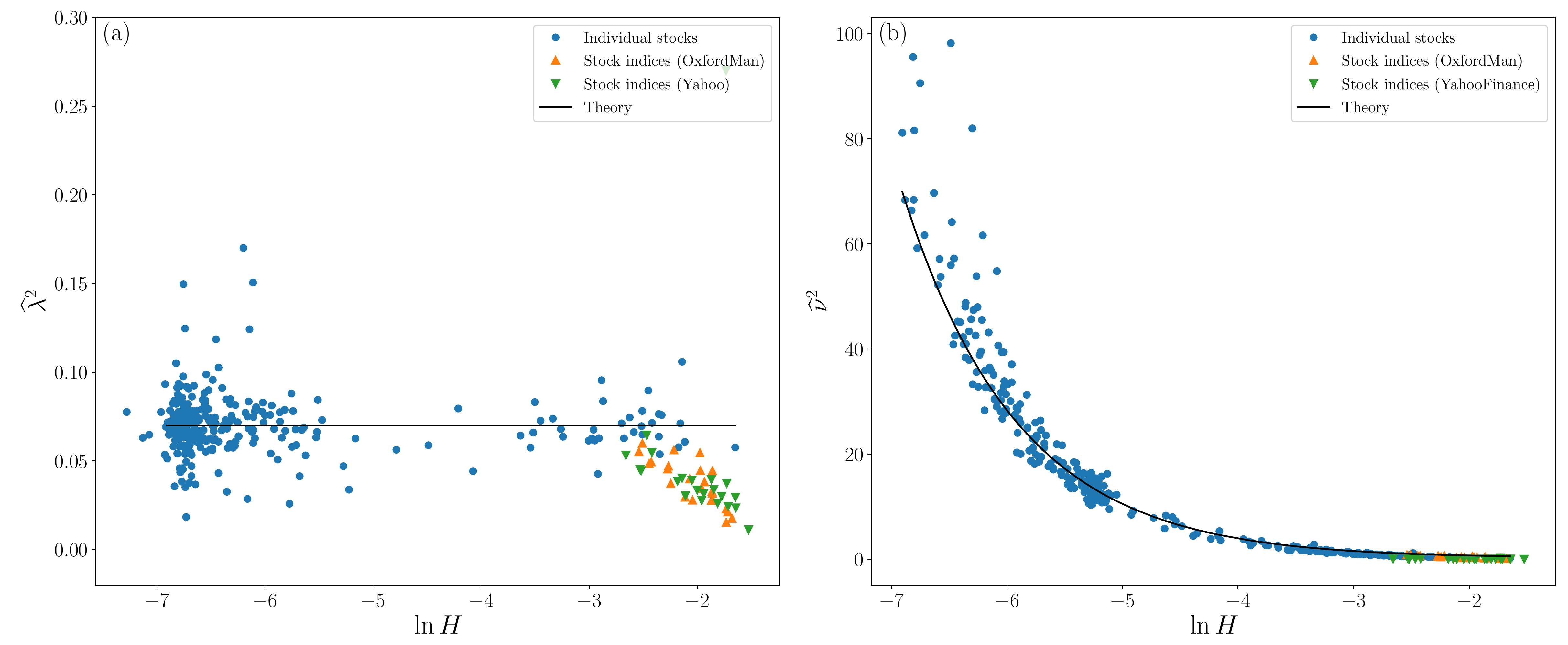}
	    \vspace*{-0.5cm}
		\caption{Estimation of the intermittency and variance parameters as a function of the estimated Hurst exponent.
			(a) Estimated intermittency coefficient $\widehat{\lambda^2}$ as a function of the logarithm of the estimated Hurst exponent $H$. The solid line represents the constant value $\lambda^2 = 0.07$ corresponding to the best fit of individual stock data.
			(b) Estimated variance coefficient $\widehat{\nu^2}$ as a function of the logarithm of the estimated Hurst exponent $H$. The solid line represents the log S-fBM expression \eqref{eq:nu_vs_H}. In (a) and (b)
			blue dots represent the individual stock data from YF database, orange up-pointing triangles represent index data from OM database while down-pointing green triangles correspond index data from YF database.}
		\label{fig:Fig_5_ln}
	\end{figure}
\end{center}
%%%%%%%%%%%%%%%%%%%%%%%%%%%%%%%%%%%%%%%%%%%%%%%%

\vspace*{-1cm}
The $\mbox{GMM}_{{\ln}M}$ estimations on all the 24 stock indices from both OM and YF databases and the 296 individual stocks from the YF database are summarized in Figures \ref{fig:Fig_5_Histo} and \ref{fig:Fig_5_ln}.
In Fig. \ref{fig:Fig_5_Histo}, we have reported the normalized histograms of the estimation ${\widehat H}$ for the Hurst exponents of the stock indices and the individual stocks of the two datasets.   
We can observe that the two distributions are quite different: while the Hurst exponents of the 
stock market indices are spread around $H \simeq 0.13$ with a rather large dispersion 
(corresponding to a rms of 0.03), the distribution of $H$ values of individual stocks is mainly peaked around a very small average value $H \simeq 0.01$ (with a rms of 0.015). It therefore clearly appears that the 
log-volatility of stock indices is much more regular than the log-volatility of individual stocks which turns out
to be well described by a multifractal model characterized by $H = 0$. Moreover, 
in agreement with the findings of \cite{vol_is_rough} (and in contrast with the results reported in \cite{rough_gmm}), Stock indices are confirmed to be well described by a "rough volatility" model with a typical value of the Hurst exponent close to $H = 0.15$. Notice that that estimation from either OM bipower variation realized volatility (orange bars) or YF Garman Klass realized volatility (green bars) provides a similar result\footnote{To be more precise, empirically the find a correlation coefficient of 0.7 between the two series of Hurst index estimates (i.e., from OM and YF data) over the 24 indices.}.

As far as the intermittency coefficient $\lambda^2$ is concerned,
we reported in Fig. \ref{fig:Fig_5_ln}(a) the estimated values $\hat \lambda^2$ for the 296 individual 
stocks (blue bullets) and the 24 stock indices (orange and green triangles for respectively OM and YF data) as a function of the logarithm of the estimated 
exponent $\hat H$. First, we can remark in both figures that OM and YF index data provide close estimations and therefore lead to the same conclusions. We can see that all the points are distributed around the value $\lambda^2 \simeq 0.07$ for stocks 
and $\lambda^2 \simeq 0.05$ for indices. 
In contrast, if one estimates the variance parameter $\nu^2$, one observes a very large dispersion of its values. Actually, as it can be checked in Fig. \ref{fig:Fig_5_ln}(b), the
data closely follow the curve $\nu^2 = \frac{0.07}{H(1-2H)}$ as represented by the solid line.  
Whether $\widehat{H}$ varies because $H$ itself is varying or because of estimation errors, it appears that $\nu^2$ is related to $\widehat{H}$ through the relationship \eqref{eq:nu_vs_H}. This suggests that the intermittency coefficient $\lambda^2$ is more likely to be the pertinent parameter to account for volatility fluctuations. Moreover, this latter quantity appears to be ``almost universal'' with a value $\lambda^2 \simeq 0.07$ for stocks and $0.05$ for indices. 
Let us remark that, because of data dispersion due to estimation errors, one can not exclude in Fig. \ref{fig:Fig_5_ln}(a), a direct relationship between $\lambda^2$ and $H$, since one can observe that $\lambda^2$ slightly decreases as the Hurst exponent increases.
A possible explanation could be that a linear combination of multifractal processes with some particular correlations in their increments or in their volatilities (the individual stock prices) may appear as a rough volatility process with an intermittency that depends on $H$ (some market index). This question will be considered in future work.

%Finally, we observed long time correlation together with $H=0.1$ for Bund future and $H=0.05$ for DAX future. 
%As far as the intermittency coefficient estimation is concerned, we observed values spread around $\lambda^2 = 0.05$ for individual stocks and stock indices while its value seems significantly greater for future data. 
%

%\begin{table}[h]
%  \centering
%  %
%  \begin{tabular}{|c|c|c|c|c|}
%    \hline
%    \multicolumn{1}{|c|}{Assets} & 
%    \multicolumn{1}{|c|}{World Indices} &
%    \multicolumn{1}{|c|}{Individual Stocks} & \multicolumn{1}{|c|}{DAX (4 mins)} &   \multicolumn{1}{|c|}{Bund (4 mins)}  \\   
%      \hline
%    $\hat H$ (GMM$_{\ln M}$) & 0.13 (0.03) & 0.045 (0.03) & 0.05 & 0.10 \\ \hline
%    $\widehat \lambda^2$ (GMM$_{\ln M}$) &  0.04 (0.01) & 0.05 (0.015) & 0.09 & 0.07       \\ \hline
%  \end{tabular}
%  %
%  \caption{Summary of $\mbox{GMM}_{{\ln}M}$ estimator performed over various assets based on daily volatility.}
%  \label{tbl_2}
%\end{table}

\section{Conclusion}
\label{sec:conc} 
We have introduced the log S-fBM, a class of log-normal ``rough'' random measures $M_{H,T} (\d t)$ that converge, when $H \to 0$, to the log-normal multifractal random measure. This model allows us to consider, within the same framework, the two popular classes of multifractal ($H = 0$) and rough volatility ($0<H < 1/2$) models. Besides the roughness exponent $H$, the model involves 3
supplementary parameters: $\sigma^2 = \frac{\E (M_{H,T}(\d t))}{\d t}$ that provides the mean value of $M_{H,T}([t,t'])$, the intermittency coefficient $\lambda^2$ which is related to the variance of $\ln M_{H,T}$ and the correlation length $T$ (also referred to as the "integral scale" in the  multifractal literature) above which the process values are independent. 
The second-order properties are studied and notably, we have computed the correlation function of $\ln M_{H,T}$ to the first order in $\lambda^2$.
By studying the self-similarity properties of $M_{H,T} (\d t)$ when one changes the correlation length $T$, it appears that one cannot estimate $T$ and $\sigma^2$ in the ``high-frequency'' estimation regime, i.e., if one observes, at a small scale $\Delta \ll T$, a single sample of $M_{H,T}$ over an interval of length $L = \mathcal{O}(T)$.

We design two efficient GMM estimation methods, $\mbox{GMM}_{M}$ and $\mbox{GMM}_{{\ln}M}$ based on the expressions of respectively $M_{H,T}$ and $\ln M_{H,T}$ correlation functions.
We provide theoretical arguments and numerical evidence showing that very much like the method introduced in \cite{bacry_kozhemyak_muzy_2013}, $\mbox{GMM}_{{\ln}M}$ provides an efficient estimation of $H$ and $\lambda^2$ even in the high-frequency asymptotic regime.

We illustrate on various numerical examples that, when $H < 1/2$,  the most pertinent parameter for 
accounting for volatility fluctuations is not, as it is always used in the rough volatility literature 
\cite{vol_is_rough,rough_gmm},  the variance parameter $\nu^2 = \frac{\lambda^2}{H(1-2H)}$, but the 
intermittency parameter $\lambda^2$. Indeed the estimation of the variance parameter is shown to 
have large fluctuations and to strongly depend on the estimation error on $H$.

Finally, when calibrating the log S-fBM model on a large set of empirical daily volatility data, we observe that stock market indices have values around $H=0.1$ (close to a rough volatility behavior)
whereas individual stocks are characterized by values of $H$ that can be very close to $0$ (close to a multifractal volatility behavior). The conclusions of section \ref{sec:gmm_estimation} concerning the Hurst estimation errors for the typical sample sizes we considered ($L \simeq 5.10^3$ corresponding to 20 years of daily data) and the agreement of the estimated values from two very different volatility estimators associated with YF and OM databases, make it unlikely that such an observation is caused by a statistical bias. Furthermore, nothing guarantees that a linear combination (indices) of multifractal processes (individual stocks) appears to be itself multifractal. It may simply appear as a "rougher" volatility process with eventually a smaller intermittency coefficient. This specific question will be addressed in future work. Finally, we pointed out that the estimations of the intermittency coefficient $\lambda^2$ are much more robust than the ones of the variance parameter $\nu^2$. Its value seems to be quite universal and spread around $\lambda^2 = 0.07$ for stocks and $\lambda^2 = 0.05$ for stock market indices in agreement with the values formerly reported for the multifractal model \cite{bacry_kozhemyak_muzy_2013}. 

In a future work, we will consider the issue of defining a faithful model for asset and option prices within the log S-fBm framework. 
To that end, the problem of introducing a specific skewness in our model will be considered along the same line as in Ref. \cite{BDM12}

%%%%%%%%%%%%%%%%%%%%%%%%%%%%%%%%%%%%%%%%%%%%%%%%%%%%%%%%%%%%%%%%
%%%%%%%%%%%%%%%%%%%%%%%%%%%%%%%%%%%%%%%%%%%%%%%%%%%%%%%%%%%%%%%%
%%%%%%%%%%%%%%%%%%%%%%%%%%%%%%%%%%%%%%%%%%%%%%%%%%%%%%%%%%%%%%%%
\appendix
\section{Appendix}
\subsection{Construction of the S-fBM process}
\label{Appendix_geo_construction}
In this Appendix, we explain in every details how the S-fBM process $\omega_{H,T}(t)$ is defined. It depends on three parameters : 
\begin{itemize}
\item  the (Hurst) parameter $H \in ]0,1[$, 
\item the decorrelation time scale $T>0$ 
\item and the coefficient $\lambda^2>0$ which is linked to the variance parameter $\nu^2$ by
$$\nu^2 = \frac{\lambda^2}{H(1-2H)}.$$
This parameter will be referred to as the intermittency parameter since it controls the intensity of intermittent ``bursts'' observed in $M_{H,T}$ and it is the name given to that quantity in the framework of MRM. 
\end{itemize}

\begin{figure}[H]\label{fig_def_C}
\centering
\includegraphics[width=0.6\textwidth]{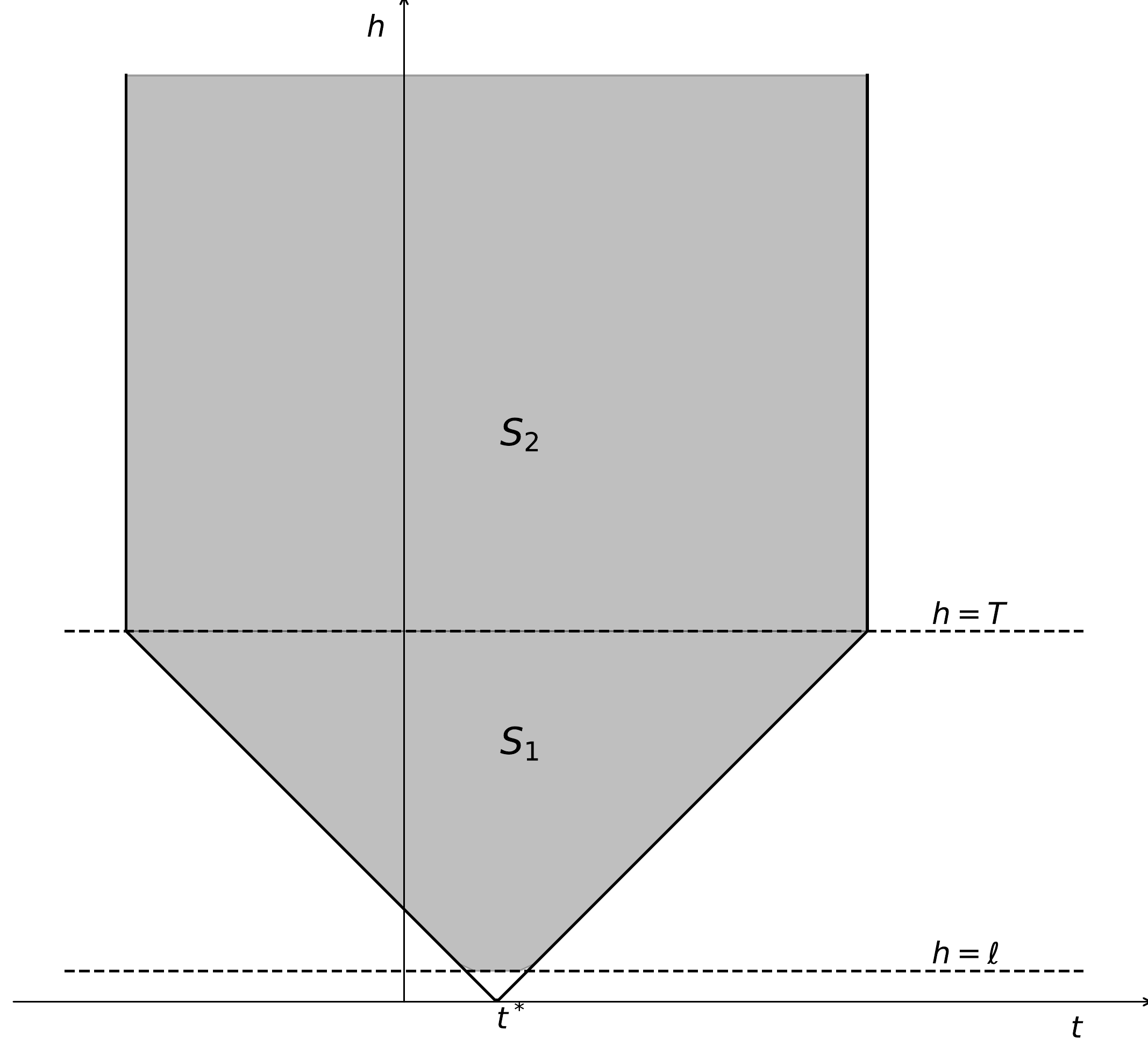}
\caption{Definition of time-scale domain $C_{\ell,T}(t)$}
\label{fig:cone}
\end{figure}

\vskip .2cm
\noindent
{\bf Construction of the S-fBM process $\omega_{H,T}(t)$} \\
In the upper half-plane $(t,h) \in {\cal S} = \mathbb{R} \times \mathbb{R^\star}$, we first consider the area $C_{\ell,T}(t^*)$ illustrated in Fig. \ref{fig:cone} which is defined as:
\begin{equation}
C_{\ell,T}(t^*) = \{(t,s)|h>\ell,|t-t^*| < \frac{1}{2} \min(h,T) \}.
\end{equation}

\noindent
For $\ell=0$, we will use the notation $C_{T}(t^*) = C_{0,T}(t^*)$.

\noindent
We then consider in $\cal S$ a non homogeneous Gaussian white noise $dG_{H}(t,h)$ of variance:
\begin{equation}
\label{vardG}
\d p_{H}(t,h) = \E \left(\d G_H(t,h)^2 \right) = \lambda^2 h^{2H-2} \d h \d t.
\end{equation}

\noindent
We will see below that $H$ is the analog of the Hurst parameter of the fBM process. 

\noindent
We then define the Gaussian process $\omega_{H,T}(t)$ as:
\begin{equation}
\label{def:omega_H}
\omega_{H,T}(t)= \mu_{H,T}(t) + \int_{C_{T}(t)} \! \! \! \! \! \d G_{H},
\end{equation}
where $\mu_{H,T}(t)$ is a normalising constant such that
\begin{equation}
\E[e^{\omega_{H,T}(t)}] = 1.
\end{equation}

% \noindent
% If $\omega_{\ell,H,T}(t) = \mu_{\ell,H,T} + \int_{C_{\ell,T}} dG_H$ with $\mu_{\ell,H,T}$ such that 
% $\E[e^{\omega_{\ell,H,T}(t)}] = 1$, then one has trivially, 
% almost surely, $\forall t$,
% \begin{equation}
% \omega_{H,T}(t)= \lim_{\ell\to 0} \omega_{\ell,H,T}(t) \; .
% \end{equation}

% When $0<H<1$, the log S-fBM measure of the interval $I$, is the random measure:

% \begin{equation}
% M_{H,T}(I)= \int_I e^{\omega_{H,T}(s)} \d s
% \end{equation}

\begin{figure}[H] \label{fig_def_cov}
\centering
\includegraphics[width=0.7\textwidth]{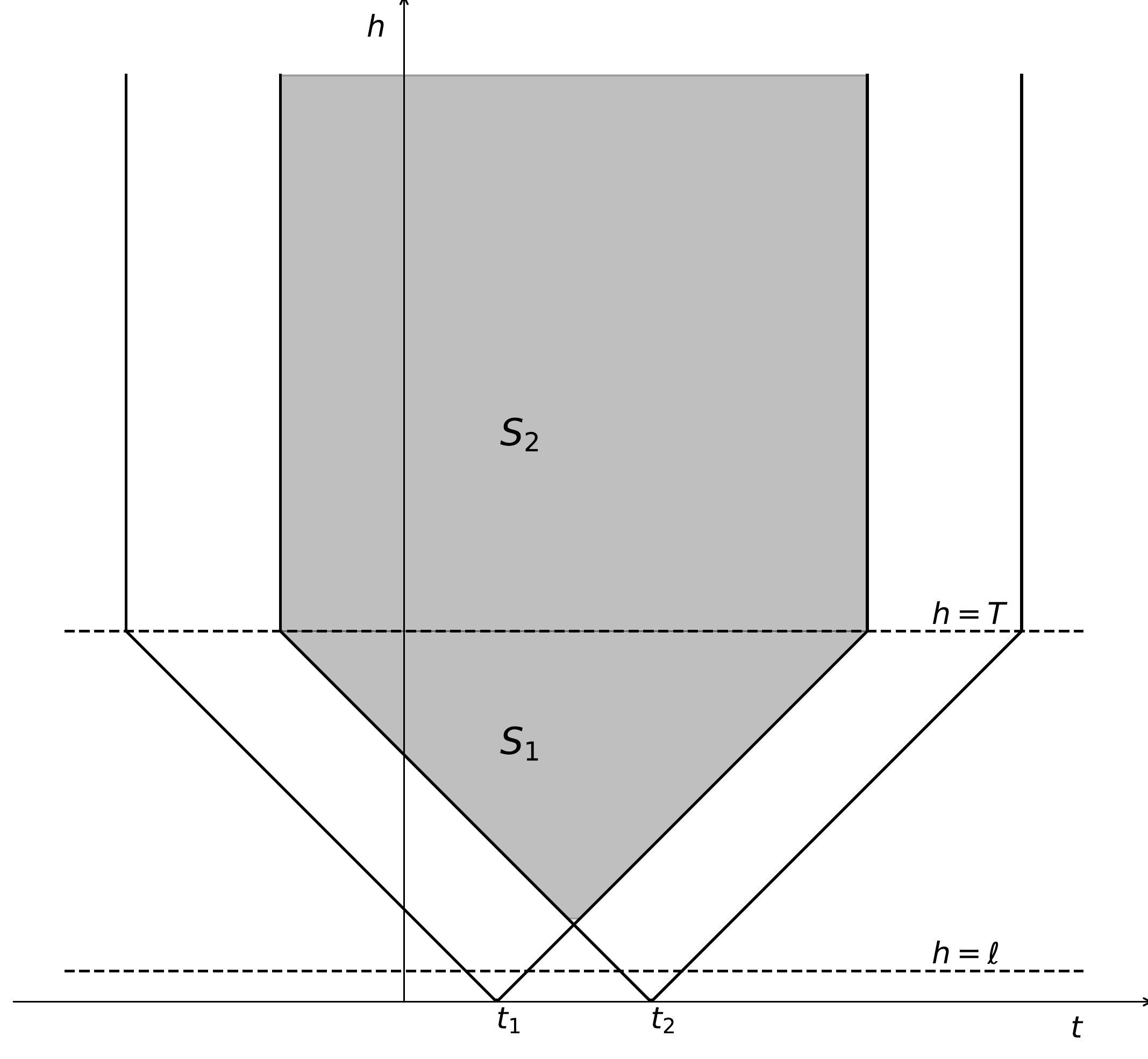}
\caption{The overlapping area}
\label{Fig:Overlap}
\end{figure}

\noindent
{\bf Covariance function of the S-fBM process $\omega_{H,T}(t)$} \\
As a Gaussian process, the S-fBM is mainly characterised by its covariance function. 
This covariance can be directly calculated as the variance of integral of the random measure $\d G_H(t,h)$ on the overlapping area of $C_{T}(t_1)$ and $C_{T}(t_2)$ displayed in Fig. \ref{Fig:Overlap}:

\begin{equation}
\Cov(\omega_{H,T}(t_1), \omega_{H,T}(t_2))= \Var \left( \int_{C_{\ell,T}(t_1) \cap C_{\ell,T}(t_2)}  \mspace{-80mu}  \d G_{H} \mspace{50mu} \right).
\end{equation}

\noindent
Let us assume, without loss of generality, that $t_2 > t_1$ and denote $\tau = t_2-t_1$.
When $\tau > T$, $C_{T}(t_1) \cap C_{T}(t_2) = \emptyset$ and thus 
$\Cov(\omega_{H,T}(t_1), \omega_{H,T}(t_2)) = 0$. 
For $\tau<T$, we have, using the notations of Fig. \ref{Fig:Overlap},
\begin{equation}\label{formule_cov}
\begin{aligned}
\Cov(\omega_{H,T}(t_1), \omega_{H,T}(t_2)) &= \int_{S_1 \cup S_2} \mspace{-30mu}  \d p_H(h,t)  \\
&= \int_{S_1} \d p_H(t,h) + \int_{S_2} \d p_H(h,t) .  \\
\end{aligned}
\end{equation}

\noindent
Using \eqref{vardG}, we have for the first term,
\begin{equation}
\begin{aligned}
\int_{S_1} dp_H(h,t)  
&= \lambda^2 \int_{t_2-t_1}^{T} h^{2H-2} \d h \int_{t_2-h/2}^{t_1+h/2} \d t \\ 
&= \lambda^2 \int_{\tau}^T h^{2H-2}(h-\tau) \\
&= \frac{\lambda^2}{2H} \left(T^{2H}-\tau^{2H} \right) - \frac{\lambda^2}{2H-1} \tau \left(T^{2H-1}-\tau^{2H-1}\right) .
\end{aligned}
\end{equation}

\noindent
For the second term,
\begin{equation}
\begin{aligned}
\int_{S_2} p(h,t) \d h \d t 
&= \lambda^2(T-\tau)\int_{T}^{\infty} h^{2H-2} \d h\\
&= -\frac{\lambda^2}{2H-1} (T-\tau) T^{2H-1} .
\end{aligned}
\end{equation}

%\noindent
%Suppose $\tau<<T$,
%\begin{equation}
%\Cov(\omega_{t_1}, \omega_{t_2})= Var(\omega_{t}) - \frac{2\nu^2}{2H-1}[(\frac{\tau}{2})^{2H}]
%\end{equation}
%To satisfy the condition $\mathbb{E}[e^{\omega_t }] = 1$, $\mu_{t}$ should be chosen as:
%\begin{equation}
%\mu_{t} = -2 + 2\log(l) -2\log(\frac{t}{2})
%\end{equation}

\noindent
By composing the results above,
\begin{equation}
\label{cov_sfBM}
\Cov(\omega_{H,T}(t_1), \omega_{H,T}(t_2)) =  \frac{\lambda^2}{2H(1-2H)}[T^{2H} - \tau^{2H} ] .
\end{equation}

\noindent
Similarly, if we consider a strictly positive $\ell$ and $\tau < \ell$, direct calculation shows:
\begin{equation}
\label{cov_sfBM_with_l}
\Cov(\omega_{H,T}(t_1), \omega_{H,T}(t_2)) =  \lambda^2 \left[\frac{1}{2H(1-2H)} (T^{2H} - \ell^{2H} ) +  \frac{ \ell^{2H}}{1-2H}  (1-\frac{\tau}{\ell})\right] .
\end{equation}

\subsection{Takenaka fBM and proof of Proposition \ref{prop_conv_to_fbm}}
\label{Append_proof_prop1}
Let us denote by $C(t_0)$ the full cone obtained by considering $T \to \infty$ in $C_{T}(t)$:
\begin{equation}
  C(t_0) = \{(t,h)|h>0, |t-t^*| < \frac{h}{2}  \} 
\end{equation}
and consider the domain:
\begin{equation}
  D(t) = C(t) \; \Delta \; C(0) = D_+(t) \cup  D_-(t),
\end{equation}

\noindent
where $\Delta$ stands for the symmetric difference between two sets and
$D_+(t)$, $D_-(t)$ are the two disjoint sets:
$$
 D_+(t) = C(t)-C(0) \; \; \mbox{and} \; \; D_-(t) = C(0)-C(t).
$$
Along the same line as definition \eqref{def:omega_H}, let us define the Gaussian processes:
\begin{equation}
  \omega_{\pm}(t) = \int_{D_{\pm}(t)} dG_H \; \; \mbox{and} \; \; B_H(t) = \omega_+(t)+\omega_-(t) \;.
\end{equation}
Notice that $\omega_{\pm}(0) = 0$ and therefore $B_H(0) = 0$.
It is easy to show that, after a little algebra that, for $0 \leq t_1 \leq t_2$:
\begin{eqnarray*}
  \E \left[ \omega_-(t_1) \; \omega_-(t_2) \right] & = & \frac{\lambda^2}{2H(1-2H)} t_1^{2H} , \\
  \E \left[ \omega_+(t_1) \; \omega_+(t_2) \right]& = & \frac{\lambda^2}{2H(1-2H)} \left(t_2^{2H} - |t_2-t_1|^{2H} \right) ,\\
   \E \left[ \omega_\pm(t_1) \; \omega_\mp(t_2) \right] & = & 0 \; .
\end{eqnarray*}
It directly results that:
\begin{equation}
 \gamma(t_1,t_2) =  \E \left[ B_H(t_1) \; B_H(t_2) \right] = \frac{\nu^2}{2} \left( t_1^{2H}+t_2^{2H}-|t_1-t_2|^{2H} \right)
\end{equation}
with $\nu^2 = \frac{\lambda^2}{H(1-2H)}$. Since $B_H(0) = 0$, we see that $B_H(t)$ is nothing but a fractional Brownian motion. This construction corresponds to the 1D version of Takenaka fractional Brownian fields as discussed in \cite{Taqqu}.

\noindent
In order to prove Proposition \ref{prop_conv_to_fbm}, let us first work out 
$$R_T(t,s) = \E \left[  (\omega_{H,T}(t) - \omega_{H,T}(0)) B_H(s) \right] \; .$$
This amounts to compute the "areas" of the intersections of $D_{\pm}(s)$ with $C_{T}(t)$ and $C_{T}(0)$ respectively.
After a little algebra, one obtains, for any $0 \leq t,s \leq T$:
\begin{eqnarray*}
  R_T(t,s) & = & \gamma(t,s) + \gamma'(t,s) \; \mbox{with} \\
  \gamma'(t,s) & = & B \left( \left(\frac{T}{2}\right)^{2H}+\left(|t-s|+\frac{T}{2}\right)^{2H}-\left(t+\frac{T}{2}\right)^{2H}-\left(s+\frac{T}{2}\right)^{2H}\right)
\end{eqnarray*}
where $B$ is a positive constant depending on $H$ and $\lambda^2$.
Similarly, if $$S_T(t,s)  = \E \left[  (\omega_{H,T}(t) - \omega_{H,T}(0)) (\omega_{H,T}(s) - \omega_{H,T}(0)) \right] \; ,$$
one has if $0 \leq s,t \leq T$:
\begin{equation}
  S_T(t,s) = \gamma(t,s) \; .
\end{equation}

\noindent
Let $$Z_{H,T}(t) = \omega_{H,T}(t)-\omega_{H,T}(0)-B_H(t)$$ and 
$$
d(t,s) = \left[ \E  \left( Z_{H,T}(t) - Z_{H,T}(s) \right)^2 \right]^{1/2} \;.
$$ 
By expanding the square $\left( Z_{H,T}(t) - Z_{H,T}(s) \right)^2$ one directly obtains:

\begin{eqnarray*}
d(t,s)^2 & = & 2 \gamma(t,t) -2 R_T(t,t) + 2 \gamma(s,s) - 2 R_T(s,s) -4 \gamma(t,s) + 4 R_T(t,s) \\
         & = & -2 \gamma'(t,t) - 2 \gamma'(s,s) + 4 \gamma'(t,s)  \\
         & = & 4 B \left( \left(|t-s|+\frac{T}{2}\right)^{2H}- \left( \frac{T}{2}\right)^{2H} \right) .
\end{eqnarray*}
Therefore, $\forall t_0$, $t,s < t_0$, we have when $T \to \infty$:
\begin{equation}
  \label{dda}
  d(t,s) = C T^{H-\frac{1}{2}} |t-s|^{1/2} + o(T^{H-\frac{1}{2}}) \; .
\end{equation}
On can consider $d(t,s)$ as a metric and define $N(t_0,\varepsilon)$ as the number of 
boxes of $d-$radius $\varepsilon$ need to cover the set $[0,t_0]$.  
Let $D = \sup_{t,s \in [0,t_0]} d(t,s)$. Then, according to Dudley inequality \cite{MISHURA2018},
there exists a positive universal constant $K$ such that:
\begin{equation}
\label{Dudley}
  \E(\sup_{t \in[0,t_0]} |Z_H(t)|) \leq K \int_0^D \sqrt{\log N(t_0,\varepsilon}) \d \varepsilon .
\end{equation}
From Eq. \eqref{dda}, one has $D \sim C T^{H-\frac{1}{2}} t_0^{\frac{1}{2}}$.
Moreover, one has
$$
 N(t_0,\varepsilon) \simeq 1+ \left \lfloor \frac{t_0 T^{2H-1}C^2}{\varepsilon^2} \right \rfloor = 1+ \left \lfloor \frac{D^2}{\varepsilon^2} \right \rfloor,
$$
where $\lfloor x \rfloor$ stands for the largest integer not greater than $x$.
We then have when $\varepsilon$ is small with respect to $D$:
$$
 \log N(t_0,\varepsilon) \simeq 2 \ln(D)-2\ln(\varepsilon) .
$$
Thus
$$
\int_0^D \sqrt{\log N(t_0,\varepsilon}) d\varepsilon \sim \int_0^{D} \sqrt{2\ln(D)-2\ln(\varepsilon))} d\varepsilon = \sqrt{\frac{\pi}{2}} D ,
$$
since for $H<1/2$, $D \to 0$ when $T \to \infty$. Proposition \ref{prop_conv_to_fbm} follows directly from inequality \eqref{Dudley}.

\subsection{The case $H \to 0$: Convergence towards the MRM log-normal measure}
\label{App_MRM}
We now examine the case $H=0$ in the geometric construction above. The definition of $C_{\ell,T}$ remains unchanged and we consider the Gaussian random noise when $H = 0$, $dG_0(t,h)$ of variance:

\begin{equation}
\d p_0(t,h) = \lambda^2 h^{-2} \d h \d t
\end{equation}

\noindent
Then we define a random process $\omega_{\ell,T}$ as previously

\begin{equation}
\omega_{\ell,T}(t)= \mu_{\ell,T} + \int_{C_{\ell,T}(t)} \d G_0 .
\end{equation}

\noindent
As proven in \cite{bacry_muzy_2003}, provided $\mu_{\ell,T}$ is chosen such that 
$\E e^{\omega_{\ell,T}(t)} = 1$, when  $\ell \to 0$ we have 
\begin{equation}
 \tM_{\ell,T}(\d t) = e^{\omega_{\ell,T}(t)} \d t \xrightarrow[\ell \rightarrow 0]{w}  \tM_T(dt) ,
\end{equation}
where $\xrightarrow{w}$ stands for the weak convergence and where $ \tM_T(dt)$ is the so-called log-normal "Multifractal Random Measure" (MRW), a non trivial singular continuous random measure with exact multifractal properties \cite{Muzy_2002,bacry_muzy_2003,bacry_kozhemyak_muzy_2013}.

\noindent
In \cite{Muzy_2002,bacry_muzy_2003}, it is also shown that the covariance of 
$\omega_{\ell,T}$ reads (for $\tau \geq 0$):

\begin{equation}
\label{covMRM}
\Cov(\omega_{\ell,T}(t), \omega_{\ell,T}(t+\tau))= 
\begin{cases}
  \lambda^2 \ln\left( \frac{T}{\tau}\right)& \text{if } \ell \leq \tau \leq T \\
  \lambda^2 \left( \ln \left( \frac{T}{\ell}\right) +1-\frac{\tau}{\ell} \right) & \text{if } \tau \leq \ell \\
    0,              & \text{otherwise}
\end{cases}
\end{equation}

\noindent
We can remark that this expression of the covariance of $\omega_{\ell,T}$ in the range $\tau \geq \ell$, can be recovered from Eq. \eqref{cov_sfBM}, \eqref{cov_sfBM_with_l} when $H \to 0$. 

Let us show a strong mean square convergence of S-fBM to MRM when $H \to 0$ as claimed in Proposition \ref{prop_conv_to_fbm}.

\noindent
Since $\tM(t) = \tM([0,t])$ is regular enough, in order to establish the weak convergence we just have to prove that $\forall t$,
\begin{equation}
  \label{eq1}
  \lim_{H\to0}  \E[( M_{H,T}[0,t] - \tM_{T}[0,t])^2] = \lim_{H\to0} \lim_{\ell\to0} \E[( M_{H,T}[0,t] - \tM_{\ell,T}[0,t])^2] = 0 .
\end{equation}

\noindent
Before starting, it is useful to calculate the covariance between $\omega_{H,T}=\omega_{\ell=0,H,T}$ and $\omega_{\ell,T}= \omega_{\ell,H=0,T}$. Following similar computation as in Appendix \ref{Appendix_geo_construction}, 
  
\begin{equation}\label{eq_corr_sfBM_MRW}
  \Cov(\omega_{H,T}(t_1), \omega_{\ell,T}(t_2))= 
  \begin{cases}
    \frac{\lambda^2}{H(1-H)}(T^{H} - \tau^{H})   & \text{if } \ell \leq \tau \leq T \\
    \lambda^2 \Big( \frac{1}{H(1-H)} (T^{H} - \ell^{H} ) +  \frac{ \ell^{H}}{1-H}  (1-\frac{\tau}{\ell}) \Big) & \text{if } \tau < \ell \\
    0             & \text{otherwise}
  \end{cases}
\end{equation}
By expanding the square in Eq. \eqref{eq1}, we have: 
\begin{multline}
   \E[(M_{H,T}[0,t] - \tM_{\ell,T}[0,t])^2] = \! \! \int_{0}^{t} \!  \! \! \int_{0}^{t}  \left(  e^{\Cov[\omega_{H,T}(u),\omega_{H,T}(v)]} + e^{\Cov[\omega_{\ell,T}(u),\omega_{\ell,T}(v)]} \right. \\  \left. -2  e^{\Cov[\omega_{H,T}(u),\omega_{\ell,T}(v)]}  \right) \d u \d v.
\end{multline}
Since, for a symmetric function $f$, one has:
$$
  \int_0^t \! \! \int_0^t f(u-v) \; \d u \d v = 2 \int_0^t (t-z) f(z) \d z.
$$
Then the previous expression becomes:
\begin{multline}
  \E[(M_{H,T}[0,t] - \tM_{\ell,T}[0,t])^2] = 2 \int_{0}^{t} \d z (t-z)  \left(  e^{\Cov[\omega_{H,T}(0),\omega_{H,T}(z)]} + \right. \\  \left. e^{\Cov[\omega_{\ell,T}(0),\omega_{\ell,T}(z)]} 
  -2  e^{\Cov[\omega_{H,T}(0),\omega_{\ell,T}(z)]}  \right) .
\end{multline}
Let us split this integral as a sum of two integrals, $I_1$ and $I_2$ according to whether one considers the integration domains $z > \ell$ and $z \leq \ell$ respectively.
In the first case, by replacing the covariance by their expressions,  one has:
\begin{equation}
I_1 = \int_\ell^t  (t-z) \left(e^{\frac{\lambda^2}{2H(1-2H)}(T^{2H} - z^{2H})} + \left( \frac{T}{z} \right)^{\lambda^2} -2e^{\frac{\lambda^2}{H(1-H)}(T^H - z^{H})} \right) \d z .
\end{equation}
Since $\lambda^2 < 1$, one can the safely take $\ell \to 0$ in the lower integral bound and then, thanks to dominated convergence theorem, observe that $I_1$ converges to 0 when $H \to 0$ since the expression inside the integral vanishes
in this limit.
The second integral, when $z \leq \ell$ is:
\begin{equation}
I_2 = \int_{0}^{\ell} (t-z) \left(
e^{\frac{\lambda^2}{2H(1-2H)}(T^{2H} - \tau^{2H})}  + 
e^{\lambda^2 (\frac{T}{\ell} + 1-\frac{z}{\ell} )} -
2e^{\lambda^2 \Big( \frac{T^{H} - \ell^{H}}{H(1-H)}  +  \frac{ \ell^{H}}{1-H}  (1-\frac{z}{\ell}) \Big)} \right) \d z.
\end{equation}
For $0 \leq z \leq \ell$,  the first and last terms inside the integral can be bounded by a constant that does not depend on $\ell$ while the second term can be bounded by $C \ell^{-\lambda^2}$. Therefore we can see that, if $\lambda^2 < 1$, $I_2 \to 0$ when $\ell \to 0$.    
This concludes the proof.

\subsection{Proof of Eqs. \eqref{eq:corrM1} and \eqref{eq:corrM2}}      
\label{App_Corr_Sfbm}
Let us compute the analytical expression of $C_M(\tau,\Delta) = \E [M_{H,T,\Delta}(t) M_{H,T,\Delta}(t+\tau) ]$
and establish expressions \eqref{eq:corrM1} and \eqref{eq:corrM2}.
For that purpose, let us first remark that from the definition \eqref{def:IV} of $M_{H,T,\Delta}(t)$ and from the 
expression \eqref{S-fBM_cov} of the covariance of $\omega_{H,T}(t)$,
we have (when $\tau < T$):
\begin{equation}
  \label{eq:c1}
  C_M(\tau,\Delta) = K_1 \int_0^\Delta \d u \int_{\tau}^{\tau+\Delta} \d v \; e^{- K_2 |u-v|^{2H}} 
\end{equation}
with $K_1 = e^{K_2 T^{2H}}$ and $K_2 = \frac{\lambda^2}{2H(1-2H)}$.

\noindent
Moreover, let us prove that, if $f(z)$ is a symmetric function, then
\begin{equation}
  \label{cov_2D1D}
   \int_0^\Delta \d u \int_{\tau}^{\tau+\Delta} \d v f(u-v) = \int_0^\Delta \d z (\Delta-z)\Big(f(z+\tau)+f(z-\tau)\Big) .
\end{equation}
Indeed, as shown in \cite{RBM19}, we have, when $\tau = 0$:
$$
\int_0^\Delta \d u \int_{0}^{\Delta} \d v f(u-v) = 2 \int_0^\Delta \d z (\Delta-z) f(z) .
$$
In the l.h.s. of \eqref{cov_2D1D}, let us set $v' = v-\tau$ and use respectively symmetry argument and 
previous expression to obtain
\begin{eqnarray*}
  l.h.s. & = & \int_0^\Delta \d u \int_{0}^{\Delta}  \d v  f(\tau +u-v) \\
         & = &  \int_0^\Delta \d u \int_{0}^{u}  \d v  f(\tau +|u-v|) + \int_0^\Delta \d u \int_{u}^{\Delta}  \d v  f(\tau -|u-v|) \\
         & = & \frac{1}{2} \left(\int_0^\Delta \d u \int_{0}^{\Delta}  \d v  f(\tau +|u-v|) + \int_0^\Delta \d u \int_{0}^{\Delta}  \d v  f(\tau -|u-v|) \right) \\
         & = & \int_0^\Delta \d z (\Delta-z)\Big(f(z+\tau)+f(z-\tau)\Big) .
\end{eqnarray*}

\noindent
By using \eqref{cov_2D1D} in \eqref{eq:c1}, we have:
\begin{eqnarray*}
  C_M(\tau,\Delta) & = & K_1 \int_0^\Delta \d z \; (\Delta-z) \; \Big( e^{- K_2 |\tau+z|^{2H}} e^{- K_2 |\tau-z|^{2H}} \Big) \\ 
   & = &  K_1 \int_\tau^{\tau+\Delta} \d z \; (\Delta+\tau-z) e^{- K_2 |z|^{2H}} + K_1 \int_{\tau-\Delta}^\tau \d z \; ,(\Delta-\tau+z)e^{- K_2 |z|^{2H}} \\
   & = & F(\tau+\Delta)+F(\tau-\Delta) - 2 F(\tau) ,
\end{eqnarray*} 
where we have denoted
$$
F(x) = K_1 \Big( x \int_0^x \d z\; e^{- K_2 |z|^{2H}} - \int_0^x \d z \; z e^{-K_2 |z|^{2H}} \Big) \; .
$$
If one considers the lower-incomplete Gamma function $\biggamma(a,z)$,  
$$
\biggamma(a,x) = \int_0^x t^{a-1} e^{-t} \d t 
$$
and makes the change of variable $t = K_2 |z|^{2H}$ in previous integrals, one obtains the following exact expression for $F(x)$:
$$
F(x) = \frac{K_1}{2H} \Big( \frac{x}{K_2^{\frac{1}{2H}}}  \biggamma(\frac{1}{2H},K_2 x^{2H})  - \frac{1}{K_2^{\frac{1}{H}}} \biggamma(\frac{1}{H},K_2 x^{2H}) \Big),
$$
which corresponds to Eq. \eqref{eq:corrM2}.
When $H = 0$, i.e. for $\tM_{T}$, one can show that the 
former expression reduces to:
$$
F(x) = \frac{T^{\lambda^2}}{(2-\lambda^2)(1-\lambda^2)} x^{2-\lambda^2} .
$$

\subsection{Proof of proposition \ref{prop_highfreq}}
\label{app:hf-regime}

In this section we provide a proof of Proposition \ref{prop_highfreq} based on small intermittency approximation of Proposition \ref{main_approx_prop}. Let $\Delta =1$ and $N = \frac{L}{\Delta} = L$ be the number of samples $M_{H,T,1}$ in the interval $[0,L]$. We will suppose that $\L \to \infty$ with $T = C L$, so that we are in the high frequency regime.
\noindent
Let us consider the empirical mean:
\begin{equation}
  \label{eq:defhatmu}
\widehat{\mu}_{N} =  \frac{1}{N} \sum_{k=1}^N  \ln M_{H,T,\Delta=1}(k)
\end{equation}
and define the ``centered'' random variable: 
\begin{equation}
\label{eq:defZ}
  Z(k) =  \ln M_{H,T,1}(k)-\widehat{\mu}_N \; .
\end{equation}
If $C_Z(k) = \Cov\left[Z(j),Z(j+k)\right]$, one has obviously:
\begin{equation}
  \label{c1}
  C_Z(k) = C_{\ln M}(1,k) - \Var \left[ \widehat{\mu}_{N} \right]\; .
\end{equation}
%From previous relationship, we have
%\begin{equation}
%\label{c1}
% C'_Z(k,\Delta) = K_0+ C_Z'(k,\Delta) - \Var \left[ \widehat{\mu}_{N} \right]
%\end{equation}
One can use Proposition \ref{main_approx_prop} to compute, to the first order in $\lambda^2$, all terms in Eq. \eqref{c1}. Indeed, the expression of $C_{\ln M}(k)$ is provided by Proposition \ref{CZ_prop} (Eq. \eqref{eq:C_Z}) and order to compute $\Var \left[  \widehat{\mu}_{\Delta, N} \right]$, one can use Prop. \ref{main_approx_prop} to show that, to the first order in $\lambda^2$,
$$
 \Var \left[ \widehat{\mu}_{N} \right] = \Var \left[ \ln M_{H,T,\Delta = L} \right]
$$
and therefore, from expression \eqref{eq:C_Z}, one has:
\begin{equation}
  \Var \left[  \widehat{\mu}_{\Delta, N} \right]  =  \frac{\lambda^2}{2H(1-2H)} \Big(T^{2H} - \frac{L^{2H}}{(2H+1)(H+1)} \Big) \; .
\end{equation}
It thus results that:
\begin{equation}
  \label{eq:cz}
  C_Z(k) = \frac{\lambda^2}{2H(1-2H)} \left(\frac{L^{2H}}{(1+2H)(1+H)}  -  \frac{|k+1|^{2H+2}+|k -1|^{2H+2}-2|k|^{2H+2}}{(2H+1)(2H+2)} \right) .
\end{equation}

\noindent
Let us consider the empirical covariance:
\begin{equation}
\widehat{C}_{\ln M}(1,k) =  N^{-1} \sum_{k=1}^{N-k} \Big(\ln M_{H,T,1}(j) - \widehat{\mu}_{N} \Big)\Big(\ln M_{H,T,1}((j+k)) - \widehat{\mu}_{N} \Big),\
\end{equation}
Since one has
$\E \left[ D(n) \right] = \E \left[ \widehat{C}_{\ln M}(1,n) - \widehat{C}_{\ln M}(1,0) \right] = C_Z(k)-C_Z(0) = \widetilde{D}(n)$
(as defined in Eq. \eqref{expr1}), in order to prove Eq. \eqref{conv1}, it is sufficient to show that
$$
 \lim_{N \to \infty} \Var \left[ D(n) \right] = 0
$$
To that end, remark that, from the definition of $D(n)$,
\begin{eqnarray}
\Var \left[ D(n) \right] & = & \!\! N^{-2} \sum_{i=1}^{N-n} \sum_{j=1}^{N-n} \Cov \left [ Z(i)Y(i),Z(j)Y(j)  \right] \\
  & = &  \!\! N^{-2} \sum_{i=1}^{N-n} \Var \left [ Z(i)Y(i) \right] + 2 N^{-2} \sum_{i=1}^{N-n} \sum_{j=i+1}^{N-n} \Cov \left [ Z(i)Y(i),Z(j)Y(j)  \right] \label{c2} 
\end{eqnarray}
where we have denoted
$$
 Y(i) = Z(i+n)-Z(i).
$$
From proposition \ref{main_approx_prop}, because $\Omega_{T,H,\Delta}(t)$ is a Gaussian process, we have, to the first order in $\lambda^2$,
\begin{eqnarray*}
 \Cov \left [ Z(i)Y(i),Z(j)Y(j)  \right]  & = & \Cov \left(Z(i),Z(j)\right) \Cov \left(Y(i),Y(j)\right) + \Cov \left(Z(i),Y(j)\right) \Cov \left(Z(j),Y(j)\right)  \\ & = & C_Z(j-i) \Big(2C_Z(j-i)-C_Z(i-j+n)-C_Z(j-i-n) \Big) \\ & +& \Big(C_Z(j+n-i)-C_Z(j-i) \Big) \Big(C_Z(j-i-n)-C_Z(j-i) \Big)  \; .
\end{eqnarray*}

\noindent
Thereby, from the expression \eqref{eq:cz} of $C_Z(k)$, after a little algebra, one can show that there exists a constant $C$ such that
$$
\Cov \left [ Z(i)Y(i),Z(j)Y(j)  \right] \leq  C N^{2H} (1+|i-j|)^{2H-2} \; .
$$

\noindent
Then, Eq. \eqref{c2} gives:

\begin{equation}
 \Var \left[ D(n) \right]  \leq  C N^{2H-1} +  2 C N^{2H-2} \sum_{i=1}^N \int_{i}^N x^{2H-2} dx \leq C' N^{2H-1}                  
\end{equation}
and thus, if $H < \frac{1}{2}$,
$$
\lim_{N \to \infty} \Var \left[ D(n) \right] = 0. \; 
$$
This concludes the proof of Eq. \eqref{conv1}.

\section*{Acknowledgement}
This research is partially supported by the Agence Nationale de la Recherche as part of the “Investissements d’avenir” program (reference ANR-19-P3IA-0001; PRAIRIE 3IA Institute).

% \bibliographystyle{plain}
% \bibliography{reference}

\end{document}